\documentclass{ifacconf}
\usepackage{amsmath}
\usepackage{amssymb}
\usepackage{mathabx}
\usepackage{tikz}
\usepackage{comment}
\usepackage{xspace}
\usepackage{subfigure}
\usepackage{graphicx}      
\usepackage{natbib}        
\usetikzlibrary{arrows,automata,positioning,fit,calc,shapes}
\usepackage{amsfonts,amssymb, bm, bbm, subfigure}

\newcommand{\MATLAB}{\textsc{Matlab}\xspace}

\newcommand{\ba}{\begin{array}}
\newcommand{\ea}{\end{array}}

\newcommand{\be}{\begin{equation}}
\newcommand{\ee}{\end{equation}}

\newcommand{\mc}{\mathcal}

\newcommand{\1}{\mathbbm{1}}

\newcommand{\R}{\mathbb{R}}
\newcommand{\N}{\mathbb{N}}

\begin{document}
\begin{frontmatter}

\title{On a Centrality Maximization Game}

\thanks[footnoteinfo]{Giacomo Como is also with the Department of Automatic Control, Lund University, Sweden. This work was partially supported by MIUR grant Dipartimenti di Eccellenza 2018--2022 [CUP: E11G18000350001], the Swedish Research Council, and by the Compagnia di San Paolo.}

\author[First]{Maria Castaldo} 
\author[Second]{Costanza Catalano} 
\author[Second]{Giacomo Como}
\author[Second]{Fabio Fagnani}

\address[First]{Univ.\ Grenoble Alpes, CNRS, Inria, Grenoble INP, GIPSA-lab, F-38000 Grenoble, France\\
(e-mail: Maria.Castaldo@gipsa-lab.grenoble-inp.fr).}

\address[Second]{Department of Mathematical Sciences ``G.L.~Lagrange'', Politecnico di Torino, Corso Duca degli Abruzzi 24, 10129 Torino, Italy\\ 
(e-mail: \{costanza.catalano,giacomo.como,fabio.fagnani\}@polito.it).}


\begin{abstract}
The Bonacich centrality is a well-known measure of the relative importance of nodes in a network. This notion is, for example, at the core of Google's PageRank algorithm. In this paper we study a network formation game where each player corresponds to a node in the network to be formed and can decide how to rewire his $ m $ out-links aiming at maximizing his own Bonacich centrality, which is his utility function. We study the Nash equilibria (NE) and the best response dynamics of this game and we provide a complete classification of the set of NE when $m=1$ and a fairly complete classification of the NE when $ m=2 $. 
Our analysis shows that the centrality maximization performed by each node tends to create undirected and disconnected or loosely connected networks, namely 2-cliques for $m=1$ and rings or a special ``Butterfly''-shaped graph when $m=2$. Our results build on locality property of the best response function in such game that we formalize and prove in the paper.
\end{abstract}


\end{frontmatter}

\section{Introduction}
The notion of \emph{centrality} aims at capturing the importance of a node in a network. This concept arises and finds application in many fields; for example, it selects the nodes in a network that have more chances to lead to cascade effects if hit by a shock (\cite{Ballester06}), or it identifies the nodes that have more influence in the opinion formation and diffusion in a social network (\cite{Kempe15}), in order to possibly perform optimal targeting interventions (\cite{Galeotti09}, \cite{Galeotti17}).
In the literature different definitions of centrality can be found, such as the \emph{degree} centrality or the \emph{eigenvalue} centrality (see for references \cite{latora17}, Section 2.3); in this paper we focus on the so-called \emph{Bonacich} centrality measure, introduced in a seminal paper by the American sociologist \cite{PB:87}. 
Formally, the Bonacich centrality $ \pi_i $ of a node $ i $ in a directed unweighted network is defined as 
\begin{equation}\label{centrality}
\pi_i=\beta\sum\limits_{j\in N_i^{-}}\frac{\pi_j}{d_j}+(1-\beta)\eta_i\, ,
\end{equation}
where $N_i^{-}$ is the in-neighborhood of node $i$ in the network, $d_j$ is the out-degree of node $ j $,  $\eta_i$ can be interpreted as the a-priori centrality of $ i $ (possibly the same for all nodes), and $\beta\in (0,1)$ is some fixed parameter.  Notice that by (\ref{centrality}), the centrality of node $ i $ depends on the centrality of the nodes $ j $ linking at $ i $ (discounted by the number of their out-links) and on its intrinsic centrality. The centrality of a node is then somewhat inherited by the nodes connected to it: a node is important in the measure that important nodes have a link to it.

The Bonacich centrality have found wide applications in many contexts, as in social networks (e.g.\ representing citations among scientists), in describing Nash equilibria in networked quadratic games (\cite{Ballester06}), in production networks among firms (\cite{Acemoglu12}), and in opinion dynamics models as the Friedkin-Johnsen model (\cite{NEF-ECJ:90}). 
A famous instance of the Bonacich centrality is the so-called \emph{PageRank centrality} for web pages, introduced by  \cite{SB-LP:98}, which is at the core of modern search engines like Google. Any search query on the web leads indeed to a set of possible related web pages that are sorted and presented according to their centrality ranking by the engine. 
Due to the relevance of the PageRank centrality for the external visibility of a web page, the problem of understanding how this measure can be efficiently computed and how it can be modified by perturbing the network has recently become very popular; see for example \cite{Ishii.Tempo:2014}, \cite{Como.Fagnani:2015}. The effect on the centrality caused by adding or deleting links in the network is not obvious from the recursive definition (\ref{centrality}). It is not difficult to see that the addition of a link $ (i,j) $ always increases the centrality of the node $ j $; less clear is how it affects the centrality of node $ i $ or, possibly, of all the other nodes in the network. In a context like that of web pages, where each node can decide only where to point its out-links and the aim is to gain visibility (that is, to increase its centrality in the network), the question of how such choice modifies its centrality and what is the rewiring that can possibly optimize it, turns out to be a natural relevant question. 
A first analysis in this sense can be found in \cite{Avrachenkov06} and \cite{dekerchove08}, while \cite{Jungers10} explore computational time issues of these problems.

In this paper, we take this point of view by assuming that nodes are left free to choose their out-links and we cast the problem into a game-theoretic setting where rewards of nodes are exactly their centralities. We investigate the shapes that the network assumes when maximazing the centrality is the only driving force: we study the Nash equilibria of our game, i.e. configurations of the network in which every node is playing its optimal action, and the behavior of the best response dynamics, i.e. a discrete dynamics in which, at every time step, a random player plays an optimal action 
(see Section 3 for formal definitions). 
We can see our problem as an instance of a \emph{network formation game}, where the actions of the players (the nodes of the network) are the ones defining the underlying network structure; we refer the reader to \cite{Jackson05} for a survey on network formation games and their applications in economy and sociology. 

More in detail, we study the problem under the assumption that all nodes are allowed to place the same number $m$ of out-links. We obtain a complete classification of the Nash equilibria in the case $m=1$, and a fairly complete classification of Nash equilibria in the case $ m=2 $. Namely, we provide necessary conditions for a configuration to be a Nash equilibrium and a complete classification of strict Nash equilibria and Nash equilibria to which converges the best response dynamics (see Section 3 for formal definitions). The main message that comes from this analysis is that the centrality maximization performed by each node tends to create undirected and disconnected or loosely connected networks: the components are $2$-cliques for $m=1$, rings and a special \emph{Butterfly} graph for $m=2$.

While completing this research, we discovered that a similar game-theoretic formulation was considered in \cite{scarsini}, Section 7, where authors prove the existence of Nash equilibria for a generalized version of our game.  While \cite{scarsini} just prove the existence of Nash equilibria and show few examples, in this work we provide an almost complete characterization of Nash equilibria, which is independent and, we believe, cannot be derived from their results. 

The paper is structured as follows. In Section \ref{model} we present the game theoretical setting; Section \ref{pre} recalls classical results and definitions of game theory, while Section \ref{main} describes the main results of the paper.  All technical results and proofs are in Section \ref{sec:proofs}. Section \ref{conclusions} concludes with summary and some open problems.

\section{The model}\label{model}
In this section, we formally define the centrality maximization game and we state the problems we want to address. 

Consider a directed graph $\mc G=(\mc V,\mc E)$ where $\mc V=\{1,\dots ,n\}$ is the set of nodes and $\mc E\subseteq \mc V\times\mc V$ is the set of (directed) edges. We denote by $(i,j) \in \mc E$ a directed edge from node $i$ to node $j$. We assume throughout the paper that $\mc G$ does not contain self-loops. In- and out- neighborhoods of a node $i$ are indicated, respectively, by $N_i^-$ and $N_i$. Their cardinalities $d_i^-=|N_i^-|$ and $d_i= |N_i|$ are, respectively, the in- and the out-degree of node $i$. Under the assumption that $d_i>0$ for every $i\in\mc V$, we equip $\mc G$ with the normalized weight matrix $R$ whose entries $R_{ij}$ are defined as
\[ R_{ij}=\frac{1}{d_i}\1_{\lbrace(i,j)\in\mc E\rbrace},\]
where $ \1 $ is the characteristic function. The entry $R_{ij}$ represents the weight attributed to the link $(i,j)$.
The Bonacich centrality $ \pi=(\pi_1,\dots ,\pi_n) $ of $\mc G$ in Eq.\ (\ref{centrality}) can be more compactly written as 
\be\label{Bonacich} \pi=(1-\beta)(I-\beta R^\top)^{-1}\eta\ee
where $ I $ is the identity matrix, $\beta\in (0,1)$, $\eta\in\R^n$ is a fixed probability vector\footnote{$ v $ is
a \emph{probability} vector if $ \sum_iv_i=1 $ and $ v_i\geq 0 $ for all $ i $.} and $ R^\top $ denotes the transpose of the matrix $ R $. A direct check shows that $\pi$ is a probability vector.
Expanding (\ref{Bonacich}) in a power series, we can write the Bonacich centrality of node $i$ as
\be\label{Bonacich2}\pi_i\!=\!(1-\beta)\!\left[\eta_i+\beta \sum_j\eta_jR_{ji}+\beta^2\sum_{j,l}\eta_jR_{jl}R_{li}+\cdots \right].\ee
Interpreting $\eta$ as a vector assigning an a-priori centrality (not depending on the graph) to each node (possibly the uniform one $\eta_i=n^{-1}$ for all $ i $), formula $(\ref{Bonacich2})$ says that the Bonacich centrality of a node in the graph $\mc G$ is the discounted sum of its own centrality $\eta_i$ and of the centrality of the other nodes discounted by the weight of the paths connecting to $i$ through the constant $\beta$. Notice that the constant $(1-\beta)$ appears just to normalize $\pi$ to a probability vector.

In our setting, we start with the set of nodes $\mc V=\{1,\dots ,n\}$ and we suppose that each node $i$ is a player that assigns $m$ directed edges from $i$ to $m$ other distinct elements in $\mc V$. This construction results in a graph $\mc G$ and the Bonacich centrality of node $i$ in $ \mc G $ represents its utility. This can be thought as a classical game where 
\begin{itemize}
\item $\mc V$ is the set of players;
\item given $i\in\mc V$, the corresponding set of actions $\mc A_i$ is the family of all subsets of $\mc V\setminus\{i\}$ of cardinality $m$;
\item let $ \mathcal{A}=\prod_i\mc A_i $ and $ x=(x_1,\dots ,x_n)\in  \mathcal{A}$ a strategy profile (or \emph{configuration}). 
We define the graph $\mc G(x)=(\mc V, \mc E(x))$ where
$\mc E(x)=\{(i,j)\;|\; i\in\mc V,\; j\in x_i\}$.
Notice that by construction $\mc G(x)  $ has constant out-degree equal to $ m $.
We denote by $R(x)$ the normalized weight matrix of $\mc G(x)$\footnote{That is, $ R_{ij}(x)=m^{-1} $ if $ (i,j)\in\mc E(x) $, $ R_{ij}(x)=0 $ otherwise.}. Given $\beta\in (0,1)$ and $\eta\in\R^n$ a probability vector such that $ \eta_i>0 $ for all $ i $, we 
define the utility vector $ u(x)=(u_1(x),\dots ,u_n(x)) $ as the Bonacich centrality of $\mc G(x)$:
$$u(x)=(1-\beta)(I-\beta R(x)^\top)^{-1}\eta.$$
\end{itemize}
The game we have introduced is denoted by $ \Gamma(\mc V,\beta,\eta, m) $ to recall all the parameters entering in the construction. 

The main goal of this paper is to analyze the structure of Nash equilibria for the game $ \Gamma(\mc V,\beta,\eta, m) $ and to investigate the asymptotic behavior of its best response dynamics, which is defined in the next section. 
The game is homogeneous in the sense that we give every node the chance to place the same number $m$ of out-links in the network. A natural generalization of this problem would be to consider a different number $ m_i $ of out-links for each node $ i $; we leave this to future work.
\section{Preliminaries}\label{pre}

In this section we recall some fundamental definitions and classical results in game theory that will be used in the next sections.

Given $x\in\mc A$ and $i\in\mc V$, we adopt the usual convention to indicate with $x_{-i}\in\mc A_{-i}=\prod_{k\neq i}\mc A_k$ the vector $x$ restricted to the components in $\mc V\setminus\{i\}$ and to use the notation $x=(x_i, x_{-i})$. 

\begin{defn}\label{defn:bestResp}
Let $ i\in \mc V $ and  $x_{-i}\in\mc A_{-i}$. We define the \emph{best response set} $\mathcal{B}_i(x_{-i})$ of node $ i $ given the strategy $ x_{-i} $  as 
$$\mathcal{B}_i(x_{-i})= \text{argmax}_{x_i\in\mathcal{A}_i}u_i(x_i, x_{-i}).$$
\end{defn}
The best response set represents the set of actions of player $ i  $ that maximize his utility $ u_i $, given the strategy $ x_{-i} $ played by all the other players.
We now recall the definition of (strict) Nash Equilibria and best response dynamics.
\begin{defn}\label{defn:nashEquilibrium}
Let $x\in \mathcal{A}$ be a strategy profile. If for all $i\in\mc V$, $x_i\in \mathcal{B}_i(x_{-i})$, then  $x$ a \textit{Nash equilibrium}. 
If  for all $i\in\mc V$, $\mathcal{B}_i(x_{-i})=\lbrace x_i \rbrace$, then $x$ a \emph{strict Nash equilibrium}. We denote by $\mc N$ and $\mc N^{\text{st}}$ the set of, respectively, Nash equilibria and strict Nash equilibria. 
\end{defn}

\begin{defn}\label{defn:bestRespDynamics}
The \emph{(asynchronous) best response dynamics} is a discrete time dynamics $ Y_t $ on the state space $\mc A$ in which at every time $t\in\N$, a player $i$ is chosen uniformly at random and he revises his action by picking an element $y$ in $\mc B_i\bigl((Y_{t-1})_{-i}\bigr)$ uniformly at random.
\end{defn}

A classical result of \cite{Monderer} states that if a game is ordinal potential\footnote{A game is ordinal potential if there exists a function $ \Psi:\mathcal{A}\to \mathbb{R} $ s.t.\ $u_i(x_i,x_{-i})<u_i(x'_i,x_{-i}) \Leftrightarrow \Psi(x_i,x_{-i})<\Psi(x'_i,x_{-i}) $.}, then its best response dynamics converges in finite time with probability one to (a subset of) Nash equilibria, independently on the initial condition.
\cite{scarsini} (Proposition 7.5 and Section 7.2) proved that our game is ordinal potential, which let us formulate the following result:
\begin{prop}\label{prop:Cominetti}
The best response dynamics on $ \Gamma(\mc V,\beta,\eta, m) $ always converges in finite time with probability one to a set $ \mc N^* \subseteq \mc N$ of Nash equilibria.
\end{prop}
Typically $\mc N^*$ is a proper subset of $\mc N$. Moreover, as strict Nash equilibria are absorbing points of the best response dynamics, it holds that $\mc N^{\text{st}}\subseteq \mc N^*$; however, in general they are not equal. 
If we consider the transition graph on the configuration set $\mc A$ induced by the best response dynamics $ Y_t $, 
the set $\mc N^*$ can be described as its smallest trapping set (no edge leading out of $\mc N^*$) that is globally reachable (from every configuration in $\mc A$ there is a path leading inside $\mc N^*$).
Nash equilibria in $\mc N^*$ play a crucial role in games as they are those the best response dynamics will eventually converge to, while Nash equilibria in $\mc N\setminus\mc N^*$ will only show up in the transient behavior.

Our aim is to investigate the structure of these three sets $\mc N^{\text{st}}\subseteq \mc N^*\subseteq \mc N$ for the game $ \Gamma(\mc V,\beta,\eta, m) $ that we have introduced in the previous section.

\section{Main results}\label{main}
In this paper we focus on the case when $m=1$ and $m=2$, namely when nodes are allowed to set, respectively, one or two out-links towards other nodes. Through a characterization of the best response set $\mathcal{B}_i(x_{-i})$, we are capable of giving a full description of the three sets $\mc N^{\text{st}}$, $\mc N^*$ and $\mc N$ of Nash equilibria for $ m\!=\!1 $, and a full description of $\mc N^{\text{st}}$ and $\mc N^*$ for $m\!=\!2$, together with a necessary condition for $ \mc N $. The case $ m\!=\!2 $ presents a much more complex behavior and, for certain aspects, as complex as the general case. 

\subsection{The case of out-degree $m=1$}
In order to describe our results, it is convenient to introduce a particular family of graphs.

\begin{figure}
\begin{center}
\begin{tikzpicture}[roundnode/.style={circle,fill=black!30, inner sep=0pt, minimum size=3mm},node distance=0.7cm,line width=0.2mm, ]
 			\node[roundnode]    (q_1) {};
 			\node[roundnode]    (q_2) [ right=of q_1] {};
 			\node[roundnode]          (q_3) [right=of q_2] {};			
 			\node[roundnode](q_4) [ right=of q_3] {};
\node[roundnode]    (q_5) [ right=of q_4] {};
 			\node[roundnode]          (q_6) [right=of q_5] {}; 			
 			\node[roundnode]          (q_7) [above left=of q_1] {};
 			\node[roundnode]    (q_8) [ left=of q_1] {};
 			\node[roundnode]          (q_9) [above right=of q_1] {}; 			
 			\node[roundnode]          (q_10) [above left=of q_5] {};
 			\node[roundnode]    (q_11) [above right=of q_6] {};
 			\node[roundnode]          (q_12) [right=of q_6] {};

 \path[->] (q_1) edge [bend right]		node  {} (q_2)
 (q_2) edge [bend right]		node  {} (q_1)
 (q_3) edge [bend right]		node  {} (q_4)
 (q_4) edge [bend right]		node  {} (q_3)
 (q_5) edge [bend right]		node  {} (q_6)
 (q_6) edge [bend right]		node  {} (q_5)
 (q_7) edge	node  {} (q_1)
 (q_8) edge	node  {} (q_1)
 (q_9) edge	node  {} (q_1)
 (q_10) edge	node  {} (q_5)
 (q_11) edge	node  {} (q_6)
 (q_12) edge	node  {} (q_6)

 ;
  
\end{tikzpicture}
\end{center}
\caption{An example of a graph of type $ C_2^{3,6} $.} 
\label{fig:C1k}
\end{figure}
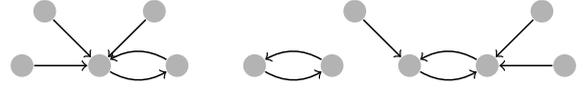

\begin{defn}
We call a \emph{$2$-clique} the complete directed graph (without self-loops) with two nodes and we indicate it by $C_2$;  we call a \textit{singleton} a node with zero in-degree. Given $l, r\in \mathbb{N}$, we define $C_2^{l,r}$ as the family of directed graph obtained by taking the disjoint union of $l$ copies of $C_2$ plus $r$ extra singletons, each of them having exactly one out-link towards a node in any of the $2$-cliques.
\end{defn}
Notice that $C_2^{l,r}$ has exactly $n=2l+r$ nodes and all nodes have out-degree equal to one. Figure \ref{fig:C1k} is an example of graph of type $C_2^{l,r}$ for $ l=3 $ and $ r=6 $.
The following theorem is our first main result for the case $m=1$.

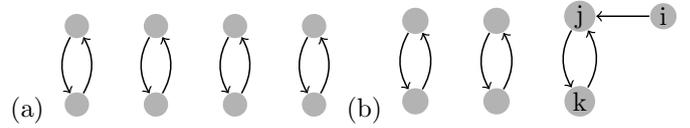
\begin{figure}
\begin{center}
(a)
\begin{tikzpicture}[roundnode/.style={circle,fill=black!30, inner sep=0pt, minimum size=3mm},node distance=0.7cm,line width=0.2mm, scale=0.9]
 			\node[roundnode]    (q_1) {\textcolor{black!30}{j}};
 			\node[roundnode]    (q_2) [ right=of q_1] {\textcolor{black!30}{j}};
 			\node[roundnode]          (q_3) [right=of q_2] {\textcolor{black!30}{j}};			
 			\node[roundnode](q_4) [ right=of q_3] {\textcolor{black!30}{j}};
			
 			\node[roundnode]          (q_7) [above=of q_1] {\textcolor{black!30}{j}};
 			\node[roundnode]    (q_8) [above=of q_2] {\textcolor{black!30}{j}};
 			\node[roundnode]          (q_9) [above=of q_3] {\textcolor{black!30}{j}}; 			
 			\node[roundnode]          (q_10) [above=of q_4] {\textcolor{black!30}{j}};

 \path[->] (q_1) edge [bend right]		node  {} (q_7)
 (q_7) edge [bend right]		node  {} (q_1)
 (q_2) edge [bend right]		node  {} (q_8)
 (q_8) edge [bend right]		node  {} (q_2)
 (q_3) edge [bend right]		node  {} (q_9)
 (q_9) edge [bend right]		node  {} (q_3)
 (q_4) edge [bend right]		node  {} (q_10)
 (q_10) edge [bend right]		node  {} (q_4)
 ;
$ \qquad \quad$  
\end{tikzpicture} (b)
\begin{tikzpicture}[roundnode/.style={circle,fill=black!30, inner sep=1pt, minimum size=2mm},node distance=0.7cm,line width=0.2mm,scale=0.9 ]
 			\node[roundnode]    (q_1) {\textcolor{black!30}{i}};
 			\node[roundnode]    (q_2) [ right=of q_1] {\textcolor{black!30}{i}};

 			\node[roundnode]          (q_6) [right=of q_2] {k}; 			
 			\node[roundnode]          (q_7) [above=of q_1] {\textcolor{black!30}{i}};
 			\node[roundnode]    (q_8) [above=of q_2] {\textcolor{black!30}{i}};

 			\node[roundnode]          (q_12) [above=of q_6] {j};
 			\node[roundnode]          (q_13) [right=of q_12] {i};
 			
 \path[->] (q_1) edge [bend right]		node  {} (q_7)
 (q_7) edge [bend right]		node  {} (q_1)
 (q_2) edge [bend right]		node  {} (q_8)
 (q_8) edge [bend right]		node  {} (q_2)
 
 (q_6) edge [bend right]		node  {} (q_12)
 (q_12) edge [bend right]		node  {} (q_6)
 (q_13) edge []		node  {} (q_12)
 ;
  
\end{tikzpicture}
\end{center}
\normalsize
\caption{(a) A graph of type $ C_2^{n/2,0} $ with $n=8$; (b) A graph of type $ C_2^{(n-1)/2,1} $ with $ n=7 $.} 
\label{fig:C1n-1/2}
\end{figure}

\begin{thm}\label{thm:nash_m=1} For any choice of $\beta$ and $\eta$, the game $ \Gamma(\mc V,\beta,\eta, 1) $  has the following properties:
\begin{enumerate}
\item the set of Nash equilibria $ \mc N $ coincides with all the configurations $x\in\mc A$ for which $\mc G(x)$ is of type $C_2^{l,r}$ with $2l+r=n$;
\item the set of strict Nash equilibria $\mc N^{st}$ is empty when $n$ is odd and it coincides with all the configurations $x\in\mc A$ for which $\mc G(x)$ is of type $C_2^{n/2,0}$ when $n$ is even.
\end{enumerate}
%
\end{thm}

Figure \ref{fig:C1n-1/2}(a) represents a strict Nash equilibrium for $ \Gamma(\mc V,\beta,\eta, 1) $ with $n\!=\!8$, while Fig.\ \ref{fig:C1n-1/2}(b) shows a nonstrict Nash equilibrium for $n\!=\!7$. The following corollary completely captures the asymptotic behavior of the best response dynamics of $ \Gamma(\mc V,\beta,\eta, 1) $; in particular it shows that the Nash equilibrium of Fig.\ \ref{fig:C1n-1/2}(b) belongs to $ \mathcal{N}^* $.


\begin{cor}\label{cor:trapping_sets}
Consider the best response dynamics for the game $ \Gamma(\mc V,\beta, \eta,1) $. For any choice of $\beta$ and $\eta$, it holds that:
\begin{itemize}
\item if $ n$ is even, the limit set $\mc N^*$ coincides with $\mc N^{\text{st}}$, namely it consists of those $x\in\mc A$ for which $\mc G(x)$ is of type $C_2^{n/2,0}$;
\item if $ n$ is odd, the limit set $\mc N^*$ coincides with those $x\in\mc A$ for which $\mc G(x)$ is of type $C_2^{(n-1)/2,1}$.
\end{itemize}
\end{cor}

Notice that when $n=2k$, the best response dynamics will eventually be absorbed in any of the $|\mc N^*|=n!2^{-k}(k!)^{-1}$ strict Nash equilibria with probability one. On the other hand, when $n=2k+1$ the best response dynamics will eventually reach the (unique) trapping set consisting of $|\mc N^*|=(n-1)n!2^{-k}(k!)^{-1}$ configurations of type $C_2^{(n-1)/2,1}$. In this case, it can be shown that the best response dynamics will keep fluctuating ergodically in the set $\mc N^*$ with uniform equilibrium probability.

\subsection{The case of out-degree $m=2$}
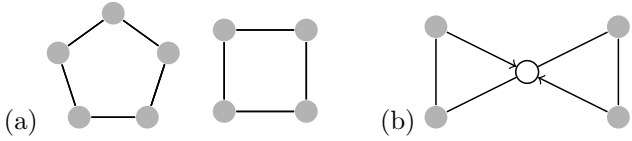
\begin{figure}
\begin{center}
(a)
\begin{tikzpicture}[roundnode/.style={circle,fill=black!30, inner sep=0pt, minimum size=3mm},node distance=1cm,line width=0.2mm, ]

\node[regular polygon, regular polygon sides=5,minimum size=1.5cm] (pent) at (0,0) {};
\node[roundnode] (q_1) at (pent.corner 1) {};
\node[roundnode] (q_2) at (pent.corner 2) {};
\node[roundnode] (q_3) at (pent.corner 3) {};
\node[roundnode] (q_4) at (pent.corner 4) {};
\node[roundnode] (q_5) at (pent.corner 5) {};

\node[regular polygon, regular polygon sides=4,minimum size=1.5cm] (quad) at (2,0) {};
\node[roundnode] (q_6) at (quad.corner 1) {};
\node[roundnode] (q_7) at (quad.corner 2) {};
\node[roundnode] (q_8) at (quad.corner 3) {};
\node[roundnode] (q_9) at (quad.corner 4) {};


\path[-] (q_1) edge	node  {} (q_2)
 (q_2) edge 		node  {} (q_1)
 (q_2) edge 		node  {} (q_3)
 (q_3) edge 		node  {} (q_2)
 (q_3) edge		node  {} (q_4)
 (q_4) edge	node  {} (q_3)
 (q_4) edge 	node  {} (q_5)
 (q_5) edge 	node  {} (q_4)
 (q_5) edge 		node  {} (q_1)
 (q_1) edge 		node  {} (q_5)
 (q_6) edge 		node  {} (q_7)
 (q_7) edge 		node  {} (q_6)
 (q_7) edge		node  {} (q_8)
 (q_8) edge		node  {} (q_7)
 (q_8) edge 	node  {} (q_9)
 (q_9) edge	node  {} (q_8)
 (q_9) edge 	node  {} (q_6)
 (q_6) edge	node  {} (q_9)
 (q_4) edge 	node  {} (q_5)
 ; 
\end{tikzpicture}
$ \qquad $(b)
\begin{tikzpicture}[roundnode/.style={circle,fill=black!30, inner sep=2pt, minimum size=3mm},node distance=0.7cm,line width=0.2mm,scale=0.6]
 			\node[roundnode]    (q_1) at (0,0) {};
 			\node[roundnode]    (q_2)  at (0,-2) {};
 			\node[roundnode,circle,draw,fill=white]          (q_3) at (2,-1) {};
 			\node[roundnode]          (q_4) at (4,0) {};	
 			\node[roundnode]          (q_5) at (4,-2) {};			
 			
 \path[->] (q_5) edge 		node  {} (q_3) 
(q_1) edge		node  {} (q_3)
 ;
 \path[-] 
 (q_3) edge		node  {} (q_2)
 (q_1) edge 		node  {} (q_2)
 (q_4) edge		node  {} (q_3)
 (q_4) edge		node  {} (q_5)
 ;
  
\end{tikzpicture}
\end{center}
\caption{(a) Example of strict Nash equilibrium for the game $ \Gamma(\mc V,\beta,\eta, 2) $ with $ n=9 $. (b) The Butterfly graph. White nodes do not have unique best response.} 
\label{fig:C2_nash}
\end{figure}

We call \emph{ring} graph an undirected graph whose vertices are arranged in a ring so that each vertex has exactly two neighbors (see for example Fig.\ \ref{fig:C2_nash}(a), where each connected component is a ring graph). The \emph{length} of a ring graph is the number of its vertices.
From now on we say that an edge $ (i,j) $ in $ \mc G $ is \emph{undirected} if also $ (j,i) $ is an edge of $ \mc G $, otherwise we call it \emph{directed}. We say that a graph is \emph{undirected} if all its edges are undirected. In figures, we represent directed edges with arrows and undirected edges with simple lines.

The first main result of this section is a complete characterization of the set of strict Nash equilibria.
\begin{thm}\label{thm:strictNash_m2}
For any choice of $\beta$ and $\eta$, the set of strict Nash equilibria $ \mc N^{st} $ of the game $ \Gamma(\mc V,\beta,\eta, 2) $ consists of all the configurations $x\in\mc A$ for which $\mc G(x)$ is the union of ring graphs.
\end{thm}

A consequence of this fact is that for any $n\geq 3$ there always exists a strict Nash equilibrium, as the ring graph of length $ n $ is always one of these.
Figure \ref{fig:C2_nash}(a) provides an example of strict Nash equilibrium with $ n=9 $.

We now investigate the structure of all Nash equilibria.
Given a Nash equilibrium $x\in \mc A$, let $\lbrace \mc G_\lambda(x)\rbrace_{\lambda=1,\dots , \Lambda}$ be the decomposition of $\mc G(x)$ in terms of its strongly connected components. The \emph{condensation graph} of $\mc G(x)$ is defined as the graph $\mc H(x)$ whose nodes are the components $\lbrace \mc G_\lambda(x)\rbrace_{\lambda}$ and where there is an edge from $\mc G_{\lambda_1}(x)$ to $ \mc G_{\lambda_2}(x)$ if there exists an edge in $\mc G(x)$ from a node in $\mc G_{\lambda_1}(x)$ to a node in $\mc G_{\lambda_2}(x)$. The condensation graph $\mc H(x)$ is directed and acyclic. The following theorem describes the topology of $\mc H(x)$ when $ x\in\mc N $, thus characterizing the structure of the Nash equilibria of the game $ \Gamma(\mc V,\beta,\eta, 2) $. We remind that a vertex is called a \emph{sink} if it has zero out-degree and it is called a \emph{source} if it has zero in-degree. 

\begin{thm}\label{thm:condensation_graph}
Let $x\in\mc A$ be a Nash equilibrium for the game $ \Gamma(\mc V,\beta,\eta, 2) $ and $\mc H(x)$ be its condensation graph on the components $\lbrace \mc G_\lambda(x)\rbrace_{\lambda}$. For any choice of $\beta$ and $\eta$, the following facts hold:
\begin{enumerate}
\item every component $\mc G_\lambda(x)$ is either a sink or a source in $\mc H(x)$ (or both if isolated);
\item every source component is either a single vertex (singleton) or a $2$-clique;
\item every sink component is either a ring or the Butterfly graph in Fig. \ref{fig:C2_nash}(b).
\end{enumerate}
\end{thm}

\begin{figure}
\begin{center}
(a)
\begin{tikzpicture}[roundnode/.style={circle,fill=black!30, inner sep=1pt, minimum size=3mm},node distance=1cm,line width=0.2mm,  ]
\node[roundnode] (q_1) at (0,-0.325){};
\node[roundnode] (q_2) at (0.75,0){};
\node[roundnode] (q_3) at (0,0.325){};

\node[roundnode] (q_10) at (3,0){};
\node[roundnode] (q_20) at (3.75,-0.325){};
\node[roundnode] (q_30) at (3.75,0.325){};

\node[roundnode,circle,draw,fill=white, inner sep=2pt, minimum size=3mm ] (q_4) at (1.4,0.5){};
\node[roundnode,circle,draw,fill=white, inner sep=2pt, minimum size=3mm ](q_5) at (2.25,0.5){};

\path[-] 
 (q_3)  edge 		    node  {} (q_2)
 (q_1)  edge   		node  {} (q_3)
 (q_1)  edge 	    	node  {} (q_2)
 ; 
 
 \path[-] 
 (q_10)  edge 		        node  {} (q_20)
 (q_20)  edge 		        node  {} (q_30)
 (q_10)  edge 	    	    node  {} (q_30)
 (q_4)  edge	        node  {} (q_5)
 ; 
 \path[->] 
 (q_4)  edge 		        node  {} (q_2)
 (q_5)  edge 		        node  {} (q_10)
 ;

\end{tikzpicture}
(b)
\begin{tikzpicture}[roundnode/.style={circle,fill=black!30, inner sep=1pt, minimum size=3mm},node distance=1cm,line width=0.2mm,  ]

\node[roundnode] (q_1) at (0,-0.325){};
\node[roundnode] (q_2) at (0.75,0){};
\node[roundnode] (q_3) at (0,0.325){};
\node[roundnode][roundnode,circle,draw,fill=white, inner sep=2pt, minimum size=3mm ]  (q_4) at (1.5,0.6){};
\node[roundnode] (q_10) at (2.25,0){};
\node[roundnode] (q_20) at (3,-0.325){};
\node[roundnode] (q_30) at (3,0.325){};

\path[-] 
 (q_3)  edge 		    node  {} (q_2)
 (q_1)  edge 	    	node  {} (q_2)
 (q_3)  edge 	    node  {} (q_1)
 ; 
 
 \path[-] 
 (q_10)  edge 		        node  {} (q_20)
 (q_20)  edge 		        node  {} (q_30)
 (q_30)  edge 	    	    node  {} (q_10)
 ; 
 \path[->] 
 (q_4)  edge []		        node  {} (q_10)
 (q_4)  edge []		        node  {} (q_2)
 ;

\end{tikzpicture}
\end{center}
\caption{Examples of nonstrict Nash equilibria for $ \Gamma(\mc V,\beta,\eta, 2) $. White nodes do not have unique best response.} 
\label{fig:C3_nonStrictNash3}
\end{figure}
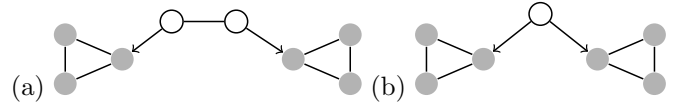

Notice that the Butterfly graph is a nonstrict Nash equilibrium as the best response of the node in the center is not unique, i.e.\ it can change action while maintaining the same utility. Figure \ref{fig:C3_nonStrictNash3} provides other two examples of nonstrict Nash equilibria: in both structures we can identify either a singleton or a 2-clique linking to rings; the nodes in white have not unique best response.


\begin{figure}
\begin{center}

\begin{tikzpicture}[roundnode/.style={circle,fill=black!30, inner sep=0.5pt, minimum size=6mm},node distance=1cm,line width=0.2mm,scale=0.5]

\node[regular polygon, regular polygon sides=6,minimum size=2cm] (esa) at (0,0) {};
\node[roundnode] (q_1) at (esa.corner 1) {2};
\node[roundnode,fill=black] (q_2) at (esa.corner 2) {\textcolor{white}{1}};
\node[roundnode,fill=black] (q_3) at (esa.corner 3) {\textcolor{white}{s}};
\node[roundnode] (q_4) at (esa.corner 4) {n-2};
\node[roundnode] (q_5) at (esa.corner 5) {n-3};
\node[roundnode] (q_6) at (esa.corner 6) {$ \dots $};
\node[roundnode] (q_7)  [above left=of q_3] {j};

\path[->] (q_7) edge 		node  {} (q_3)
(q_7) edge 		node  {} (q_2)
;

\path[-] (q_1) edge 		node  {} (q_2)
 (q_2) edge 	node  {} (q_1)
 (q_2) edge 	node  {} (q_3)
 (q_3) edge 		node  {} (q_2)
 (q_3) edge 		node  {} (q_4)
 (q_4) edge 	node  {} (q_3)
 (q_4) edge 	node  {} (q_5)
 (q_5) edge 		node  {} (q_4)
 (q_5) edge [dotted]		node  {} (q_6)
 (q_1) edge [dotted]		node  {} (q_6)
 ; 
 \node at (-3.8,-1.5) {(a)};
\end{tikzpicture}
$ \,\,\, $
\begin{tikzpicture}[roundnode/.style={circle,fill=black!30, inner sep=0.5pt, minimum size=6mm},node distance=1cm,line width=0.2mm,scale=0.5]

\node[regular polygon, regular polygon sides=6,minimum size=2cm] (esa) at (0,0) {};
\node[roundnode] (q_1) at (esa.corner 1) {2};
\node[roundnode] (q_2) at (esa.corner 2) {1};
\node[roundnode,fill=black] (q_3) at (esa.corner 3) {\textcolor{white}{s}};
\node[roundnode] (q_4) at (esa.corner 4) {n-3};
\node[roundnode] (q_5) at (esa.corner 5) {n-4};
\node[roundnode] (q_6) at (esa.corner 6) {$ \dots $};
\node[roundnode] (q_7)  [left=of q_3] {j};
\node[roundnode] (q_8)  [above left=of q_3] {k};

\path[->] (q_7) edge 		node  {} (q_3)
(q_8) edge 		node  {} (q_3)
;

\path[-] (q_1) edge 		node  {} (q_2)
(q_7) edge 		node  {} (q_8)
 (q_2) edge 	node  {} (q_1)
 (q_2) edge 	node  {} (q_3)
 (q_3) edge 		node  {} (q_2)
 (q_3) edge 		node  {} (q_4)
 (q_4) edge 	node  {} (q_3)
 (q_4) edge 	node  {} (q_5)
 (q_5) edge 		node  {} (q_4)
 (q_5) edge [dotted]		node  {} (q_6)
 (q_1) edge [dotted]		node  {} (q_6)
 ; 
 \node at (-4.5,-1.5) {(b)};
\end{tikzpicture}
\end{center}
\caption{(a) Singleton linking to two adjacent nodes in a ring. (b) 2-clique linking to a single node in a ring. Black nodes are not in best response.}
\label{fig:singleton_ring}

\end{figure}
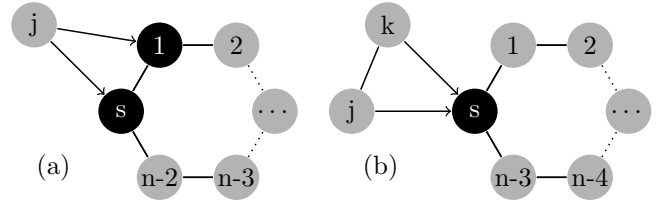

\begin{rem}
\label{remark_m2}
Not all the configurations $x\in \mc A$ that satisfy conditions (1), (2) and (3) of Theorem \ref{thm:condensation_graph} are Nash equilibria. Indeed, by direct computation it is easy to see that the following examples are not Nash equilibria:
\begin{enumerate}
    \item a singleton linking to two adjacent nodes in a ring of length greater or equal than four (see Fig.\ \ref{fig:singleton_ring}(a));
    \item a 2-clique linking to a single node in a ring of length greater or equal than four (see Fig.\ \ref{fig:singleton_ring}(b)).
\end{enumerate}
\end{rem}

We are now ready to characterize the limit set $ \mc N^* \subseteq \mc N$ for the game $ \Gamma(\mc V,\beta, \eta,2) $, i.e.\ the absorbing points of its best response dynamics.
\begin{cor}\label{cor:trapping_setsM2}
Consider the game $ \Gamma(\mc V,\beta, \eta,2) $ and let $ i$ s.t.\ $ i=n\! \mod 3 $. Then for any choice of $\beta$ and $\eta$, it holds that:
\begin{itemize}
\item if $ i= 0,1 $, the limit set $\mc N^*$ coincides with $\mc N^{\text{st}}$;
\item if $ i= 2$, the limit set $\mc N^*$ coincides with $\mc N^{\text{st}} \cup \mc G^3_b $, where $\mc G^3_b$ is the set of all graphs that are unions of rings of length three and a Butterfly graph or unions of rings of length three and a 2-clique linking to any nodes in the rings (see e.g.\ Fig.\ref{fig:C3_nonStrictNash3}(a), Fig.\ref{fig:trasitionButterfly}(b), (c)).
\end{itemize}
\end{cor}



\begin{figure}
\centering     
\includegraphics[width=0.25\textwidth]{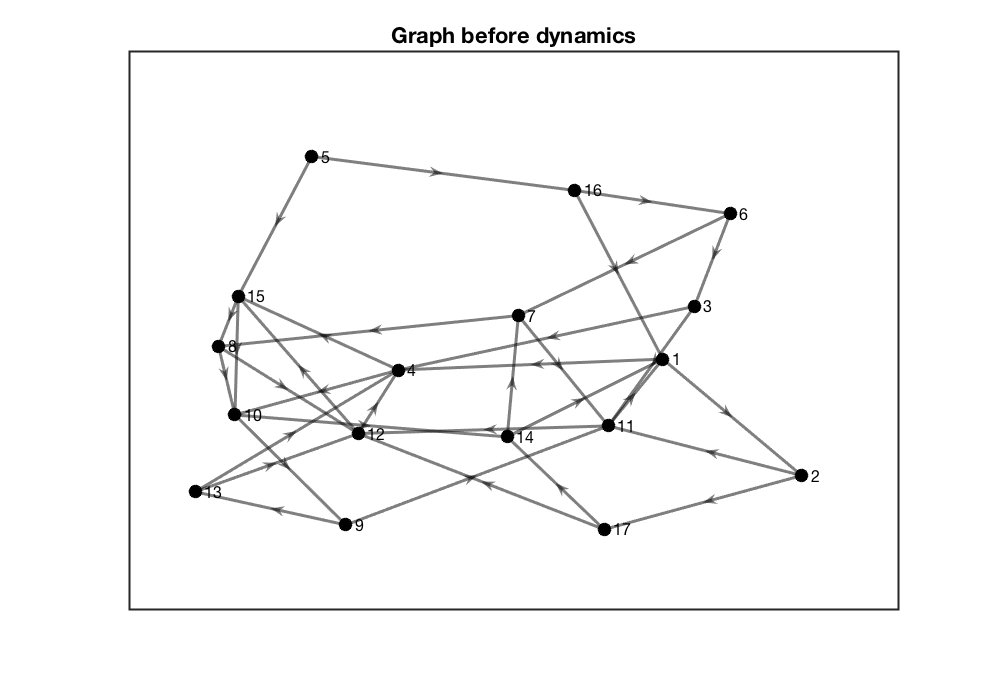}
\\
(a)\includegraphics[width=0.2\textwidth]{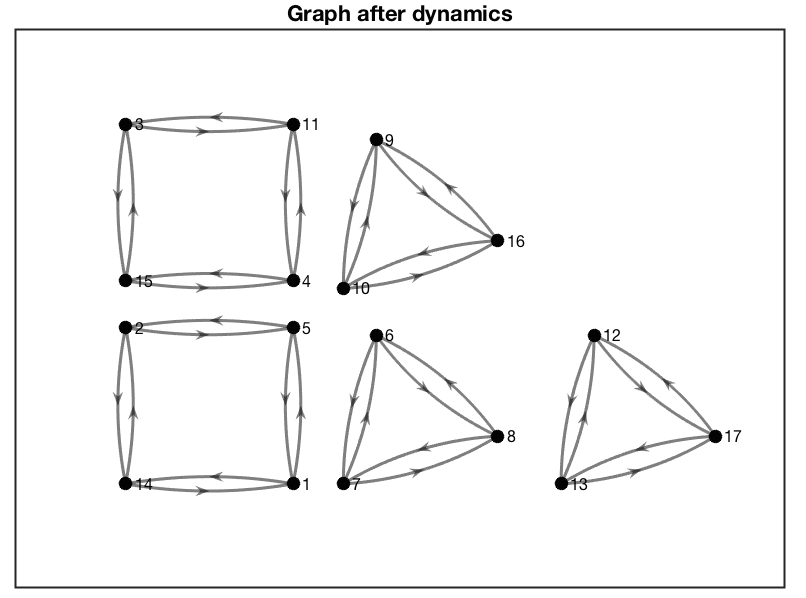}$\quad $(b)\includegraphics[width=0.2\textwidth]{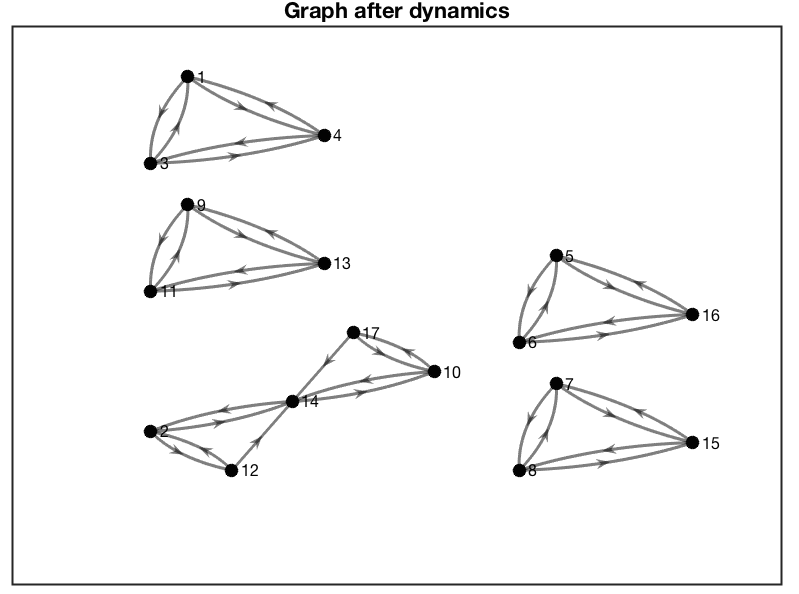}
\caption{Convergence of the best response dynamics starting from the same initial configuration to (a) a strict Nash equibrium (b) a nonstrict Nash equilibrium.}
\label{fig:dynamics}
\end{figure}

Figure \ref{fig:dynamics} shows the convergence of the best response dynamics starting from the same initial configuration to two different equilibria, namely a strict Nash equilibrium (union of rings) and a nonstrict Nash equilibrium in $ \mc G_b^3 $ (union of rings of length three and a Butterfly graph). The simulations have been done using suitable \MATLAB routines.



\section{Proofs of the results}\label{sec:proofs}
The proofs of our results are based on a probabilistic interpretation of the game in terms of Markov chains. We first recall some preliminary notions on Markov chains and we apply them to our game. Then in Subsections \ref{subsec:onelink_proofs} and \ref{subsec:twolink_proofs} we prove the results presented in the previous section respectively for the case $ m=1 $ and $ m=2 $.

A (discrete-time) Markov chain $ X_t $ on a finite state space $ \mc V=\lbrace 1,\dots ,n\rbrace $ and with transition matrix $ P\in\mathbb{R}^{n\times n} $, $ P $ stochastic\footnote{A matrix $ P $ is stochastic if each row is a probability vector.}, is a sequence of random variables $ X_1,X_2,\dots  $ with values in $ \mc V$ such that $ \mathbb{P}(X_{t+1}=i|X_1=j_1,\dots,X_t=j_t)=\mathbb{P}(X_{t+1}=i|X_{t}=j_t)=P_{j_ti} $. Given $s\in \mc V $, we define $ T_s:=\inf\lbrace t\geq 0: X_t=s \rbrace $ the \textit{hitting} time on $ s $ and $ T^+_s:=\inf\lbrace t\geq 1: X_t=s \rbrace $ the \textit{return} time to $ s $. Given $i,s\in \mc V $, we define $\tau_{i}^s:=\mathbb{E}_i[T_s] $ the \textit{expected} hitting time on $ s $ of the Markov chain $ X_t $ with initial state $ i $. It is known that if $ P$ is an irreducible matrix, then the Markov chain admits a unique invariant distribution, that is a probability vector $ \pi $ s.t.\ $ \pi=P^{\top}\pi $. The invariant distribution $ \pi $ can be written in terms of hitting times:

\begin{prop}\label{prop:hitting_times}
Let $ X_t $ be a Markov chain with finite state space $ \mc V $ and irreducible transition matrix $ P $, and let $ \pi $ be its (unique) invariant distribution. Then it holds that 
\begin{equation}\label{eq:pi}
\pi_s=\left( 1+\sum_{i\in \mc V}P_{si}\tau_i^s\right)^{-1},
\end{equation}
where the expected hitting times $\tau_i^s  $, $ i\!\in\! \mc V $, are the only family of values satisfying the following system:
\begin{equation}\label{eq:system_tau}
\begin{cases}
\tau_i^s=0 &\text{if }i=s,\\
\tau_i^s=1+\sum_{j\in \mc V}P_{ij}\tau_j^s &\text{if }i\neq s.
\end{cases}
\end{equation}

\end{prop}
\begin{pf}
Equation (\ref{eq:pi}) comes from the fact that $ \pi_s=(\mathbb{E}_s[T^+_s])^{-1}$ and $ \mathbb{E}_s[T^+_s]=1+\sum_{i\in \mc V}P_{si}\tau_i^s $, which are both standard results on Markov chains, as well as (\ref{eq:system_tau}). See for example \cite{norris_1997}.$ \qed $
\end{pf}
Manipulating (\ref{Bonacich}) and using the fact that $\mathbf{1}^\top\pi=1$ with $ \mathbf{1} $ the all-ones vector, we can see that the Bonacich centrality $\pi$ satisfies the relation 
$$\pi=(\beta R^\top+(1-\beta) \eta\mathbf{1}^\top)\pi.$$ Since  $P=\beta R+(1-\beta)\mathbf{1}\eta^\top$ is an irreducible stochastic matrix, it means that $\pi$ is the (unique) invariant distribution of the Markov chain having $ P $ as transition matrix. 
We now use this characterization in the context of our game. Given a configuration $x\in\mc A$, we write
\begin{equation}\label{eq:P(x)}
P(x)=\beta R(x)+(1-\beta)\mathbf{1}\eta^\top
\end{equation}
and we denote by $\tau_i^s(x)$ the hitting time on $ s $ of the Markov chain having $P(x)$ as transition matrix and starting from $ i $. When the configuration $ x $ is clear from the context, sometimes we write $\tau_i^s$ instead of $\tau_i^s(x)$ to ease the notation.
The utility vector $u(x)$ can be written in terms of the formula (\ref{eq:pi}) as
$u_s(x_s,x_{-s})=\left(1+\sum_{i\in \mc V}P_{si}(x)\tau_i^s(x)\right)^{-1}$.
Since  
the terms $P_{si}(x)$ only depend on $ x_s $ (the out-links from $s$), while the hitting times $\tau_i^s(x)$ only depend on $x_{-s}$, with slight abuse of notation we rewrite the utility function as
\begin{equation}\label{eq:best-times}
u_s(x_s,x_{-s})=\left(1+\sum_{i\in \mc V}P_{si}(x_{s})\tau_i^s(x_{-s})\right)^{-1}.
\end{equation}

A consequence of (\ref{eq:best-times}) is an explicit formula describing the best response set, as shown by the following remark.

\begin{rem}
\label{cor:best_response}
Consider the game $ \Gamma(\mc V,\beta,\eta, m) $, a node $ s\in \mc V $ and $ x_{-s}\in \mathcal{A}_{-s}  $. Then the best response set $ \mathcal{B}_s(x_{-s}) $ of player $ s $ when all the other players are playing the actions $ x_{-s} $ can be written as:
\begin{equation}\label{eq:best_response}
\mathcal{B}_s(x_{-s})=\underset{x_s\in\mathcal{A}_s}{ \text{argmin}}\sum_{i\in \mc V}R_{si}(x_s)\tau_i^s(x_{-s}).
\end{equation}
\end{rem}

In the following, given $ x\in \mc A $ we denote by $ N_s^{-}(x) $ the in-neighborhood of the vertex $ s $ in the graph $ \mathcal{G}(x) $, that is $ i\in  N_s^{-}(x)$ if and only if $ s\in x_i $ (or equivalently, if and only if $ R_{is}(x)>0 $). Notice that $ N_s^{-}(x) $ depends just on $ x_{-s} $ so with a slight abuse of notation we can write $ N_s^{-}(x_{-s}) $.

\subsection{The case of out-degree $ m=1 $}\label{subsec:onelink_proofs}
In order to prove Theorem \ref{thm:nash_m=1}, we need to better characterize the best response actions of a player. The first important observation is the following:

\begin{rem}\label{rem:best_response_m1}
If $ m=1 $, then for any $ s\in\mc V $ and $ x_s\in\mathcal{A}_s $ it holds that $ R_{sx_s}(x_s)=1 $ and $ R_{si}(x_s)=0 $ for all $ i\neq x_s $. Therefore (\ref{eq:best_response}) takes the form: \[ \mathcal{B}_s(x_{-s})=\text{argmin}_{i\in \mc V\setminus\lbrace s\rbrace}\tau_i^s(x_{-s}).\]
\end{rem}

The following proposition shows that the best response action of a player in the game $ \Gamma(\mc V,\beta,\eta, 1) $ takes always place in his in-neighborhood, as long as it is nonempty. 

\begin{prop}\label{prop:best_response}
Consider the game $ \Gamma(\mc V,\beta,\eta, 1) $ and let $ s\in \mc V $ and $ x_{-s}\in \mathcal{A}_{-s} $. It holds that:
\begin{enumerate}
\item If $ N_s^{-}(x_{-s})\neq \emptyset $, then   $\mathcal{B}_s(x_{-s})= N_s^{-}(x_{-s})  $; 
\item If $ N_s^{-}(x_{-s})= \emptyset $, then $ \mathcal{B}_s(x_{-s})=\mc V \setminus\lbrace s\rbrace $.
\end{enumerate}
\end{prop}

\begin{pf}
$ (1) $ Suppose that $ N_s^{-}(x_{-s})\neq \emptyset $ and let $ i,j,k\neq s $ s.t.\ $ i,j\in N_s^{-}(x_{-s})$ and $ k\notin N_s^{-}(x_{-s})$. We show that $ \tau_i^s=\tau_j^s $ and $ \tau_i^s<\tau_k^s $; by Remark \ref{rem:best_response_m1}, this implies that $\mathcal{B}_s(x_{-s})= N_s^{-}(x_{-s})  $. By Proposition \ref{prop:hitting_times}, it holds that
\begin{align*}
\tau_i^s&=1+(1-\beta)\sum_{v\in \mc V}\eta_v\tau_v^s,\,\,\,\,\, \tau_j^s=1+(1-\beta)\sum_{v\in \mc V}\eta_v\tau_v^s,\\
\tau_k^s&=1+(1-\beta)\sum_{v\in \mc V}\eta_v\tau_v^s +\beta \tau_{h}^s
\end{align*}
where $x_k=\lbrace h\rbrace $. Since $ \tau_{h}^s>0 $, it follows that $ \tau_i^s=\tau_j^s $ and $ \tau_i^s<\tau_k^s $.\\
$ (2) $ Suppose that $ N_s^{-}(x_{-s})= \emptyset $ and let $ j\neq s $. This implies that at every discrete time $ t $, the probability to arrive at node $ s $ from $ j $ is equal to $(1-\beta)\eta_s \left( 1- (1-\beta)\eta_s\right)^{t-1}  $. Therefore it holds that 
\begin{equation}\label{eq:tau_emptyinneigh}
\tau_j^s= (1-\beta)\eta_s\sum_{t=1}^{\infty}t\left( 1-(1-\beta)\eta_s \right)^{t-1},
\end{equation}
which does not depend on $ j$. We just proved that $ \tau_j^s=\tau_i^s $ for every $ i,j\neq s $, so we conclude by Remark \ref{rem:best_response_m1}. $ \quad\qed $
\end{pf}

We are now ready to prove Theorem \ref{thm:nash_m=1}.

\textbf{Proof of Theorem \ref{thm:nash_m=1}.}
$ (1) $ A configuration $ x\in \mathcal{A} $ is a Nash equilibrium iff for all $ s\in\mc V $, it holds that $ x_s\in\mathcal{B}_s(x_{-s}) $. By Proposition \ref{prop:best_response}, this happens iff for all $ s\in \mc V $ s.t. $ N^{-}_s(x_{-s})\neq \emptyset $, we have that $x_s\in N^{-}_s(x_{-s}) $, thus forming the 2-clique  $\lbrace s,x_s\rbrace $ in $ \mathcal{G}(x) $. Therefore $ x\in \mathcal{A} $ is a Nash equilibrium iff $ \mathcal{G}(x) $ is of type $ C_2^{l,r} $ where $ r $ is the number of vertices $ v$ such that $ N^{-}_v(x_{-v})= \emptyset $.\\
$ (2) $ A configuration $ x\in \mathcal{A} $ is a strict Nash equilibrium iff for all $ s\in\mc V $, it holds that $ \lbrace x_s\rbrace = \mathcal{B}_s(x_{-s}) $; by Proposition \ref{prop:best_response} this holds iff for all $ s\in\mc V $, $ N^{-}_s(x_{-s})=\lbrace x_s \rbrace$. Therefore for all $ s\in\mc V $, $ \lbrace s,x_s\rbrace $ must be a 2-clique in $ \mathcal{G}(x) $, and this is possible iff $ n $ is even and $ \mathcal{G}(x) $ is of type $ C_2^{n/2,0} $. $ \qed $

\textbf{Proof of Corollary \ref{cor:trapping_sets}.}
In view of Proposition \ref{prop:Cominetti} and Theorem \ref{thm:nash_m=1}, we just need to show that any configuration of type $ C_2^{l,r} $ will eventually converge in a best response dynamics to a configuration of type $ C_2^{n/2,0} $ when $ n $ is even and to a configuration of type $ C_2^{(n-1)/2,1} $ when $ n $ is odd. Suppose that the node $ v\in  C_2^{l,r}$ is selected in the best response dynamics; we have the following cases: 
(i) $ v $ belongs to a 2-clique and has in-degree equal to one: in this case its best response is unique so it does not change action;
(ii) $ v $ belongs to a 2-clique $ \lbrace v,w\rbrace $ and has in-degree $ >1 $: in this case by item (1) of Proposition \ref{prop:best_response}, it can change action (maintaining the same utility) by linking to some other vertex $ v_1 $ in $ N^{-}_v $. We have then two subcases: (iia) $ w $ has in-degree equal to one in $ C_2^{l,r} $ so when $ v $ changes its action, we still end up in a configuration of type $ C_2^{l,r} $; (iib) $ w $ has in-degree equal $ >1 $ in $ C_2^{l,r} $; in this case, once $ w $ is selected it \emph{has} to change action by linking back to some  $ w_1\in N^{-}_w $, $ w_1\neq v $; we hence end up in a configuration of type $ C_2^{l+1,r-2} $.
Suppose now $ v $ is one of the $ r $ vertices with zero in-degree: by item (2) of Proposition \ref{prop:best_response}, $ v $ can change action (maintaining the same utility) by linking to any other vertex $ w $ in $C_2^{l,r}  $. We have two cases:
(iii) $ w $ is a 2-clique; then we still end up in a configuration of type $ C_2^{l,r} $;
(iv) $ w $ is another vertex with zero in-degree. In this case, since now $ |N^{-}_w|>0 $, once $ w $ is selected it \emph{has} to change action by linking back to $ v $; we hence end up in a configuration of type $ C_2^{l+1,r-2} $.
We have just proved that in a best response dynamics, starting from a configuration of type $ C_2^{l,r} $ with positive probability we increase the number of two-cliques (and we can never reduce it). This implies that we will eventually converge to configurations with the maximal number of two cliques, that is $ C_2^{n/2,0} $ for $ n $ even, and $ C_2^{(n-1)/2,1} $ for $ n $ odd. 
$ \qquad\qquad\qquad\qquad\qquad\qquad\qquad\qquad\qquad\qquad\qed $

\subsection{The case of out-degree $ m=2 $}\label{subsec:twolink_proofs}
As in the case of $ m=1 $, we want to better characterize the best response set of a player. The following two lemmas will be useful for proving the subsequent Proposition \ref{prop:bestresponse_m2}, in which we show that the best response actions of a node are always towards nodes that are at most at in-distance two from it. 

\begin{lem}\label{lem:upperbound_tau}
Consider the game $ \Gamma(\mc V,\beta,\eta, 2) $, and let $ x\in \mathcal{A} $ and $ s\in \mc V $. It holds that:
\begin{enumerate}
\item for every $ i\neq s $, $ \tau_i^s(x)\leq \eta_s^{-1}(1-\beta)^{-1} $;
\item if there exists $ i\neq s $ such that $\tau_i^s(x)= \eta_s^{-1}(1-\beta)^{-1}  $, then $ N^{-}_s(x)=\emptyset $.
\end{enumerate}
\end{lem}

\begin{pf}
$ (1) $ Let $ A $ be a matrix such that for all $ i\in \mc V $, $ A_{ii}= \beta+ (1-\beta)\eta_i $ and for all $ j\neq i$, $ A_{ij}= (1-\beta)\eta_j $ . If we denote by $ \hat{\tau}_i^s $ the expected hitting time of the Markov chain $ \hat{X}_t $ with transition matrix $ A $ and initial state $ s $, by solving the system (\ref{eq:system_tau}) it is easy to see that for all $ i,k\neq s $ it holds that $ \hat{\tau}_i^s=\hat{\tau}_k^s $. This in turn implies that for every $ i\neq s $, $ \hat{\tau}_i^s= \eta_s^{-1}(1-\beta)^{-1} $. 
In $ \hat{X}_t $ the probability to jump from any node $ i $ to $ s $ is always equal to $ (1-\beta)\eta_s  $, while in the Markov chain $ X_t $ associated to our game (with transition matrix as in (\ref{eq:P(x)})) the probability to jump from any node $ i$ to $ s $ is always greater or equal than $ (1-\beta)\eta_s  $. It follows that $ \tau_i^s\leq \hat{\tau}_i^s $, so we conclude.\\
$ (2) $ Let $ i\neq s $ such that $\tau_i^s= \eta_s^{-1}(1-\beta)^{-1}  $. We first show that for every $ j\neq s $, $\tau_j^s= \eta_s^{-1}(1-\beta)^{-1}  $. Indeed, suppose by contrary that there exists $ j\neq s $ such that $\tau_j^s< \eta_s^{-1}(1-\beta)^{-1}  $. If $ a,b\in \mc V $ are the vertices such that $ x_i=\lbrace a,b\rbrace $, then by system (\ref{eq:system_tau}) it holds that \[ \tau_i^s=1+(1-\beta)\sum_{v\in \mc V}\eta_v\tau_v^s+\frac{\beta}{2}(\tau_a^s+\tau_b^s).\] In view of item $ (1) $, this implies that $ \tau_i^s< \eta_s^{-1}(1-\beta)^{-1} $, which is a contradiction; therefore $\tau_j^s= \eta_s^{-1}(1-\beta)^{-1}  $. Suppose now by contradiction that $N^{-}_s(x)\neq \emptyset $ and let $ k\in N^{-}_s(x)$ and $ a\in\mc V $ such that $ x_k=\lbrace a,s\rbrace $. By system (\ref{eq:system_tau}) it holds that
\begin{equation}\label{eq:lemma}
\tau_k^s=1+(1-\beta)\sum_{v\in \mc V}\eta_v\tau_v^s+\frac{\beta}{2}\tau_a^s.
\end{equation}
As $\tau_v^s= \eta_s^{-1}(1-\beta)^{-1}  $ for every $ v\neq s $ and $\tau_s^s=0  $, equation (\ref{eq:lemma}) implies that $ \beta=0 $, which is a contradiction and so we conclude. $\qquad\qquad\qquad\qquad\qquad\qquad\qquad\qquad\qquad \qed $
\end{pf}

The next lemma provides a different upper bound on the return times $ \tau_{i}^s(x) $ when $|N_s^-(x)| \geq 1$. 
 We denote by $ N_s^{-2}(x) $ the set $ N_s^{-}(x)\cup \lbrace  N_t^{-}(x): t\in  N_s^{-}(x) \rbrace$, that is the in-neighborhood of $ s $ in $ \mathcal{G}(x) $ at distance at most two. Notice that also $ N_s^{-2}(x) $ depends just on $ x_{-s} $ so we can write as well $ N_s^{-2}(x_{-s}) $.

\begin{lem}\label{lem:upperbound2_tau}
Consider the game $ \Gamma(\mc V,\beta,\eta, 2) $, and let $ x\in \mathcal{A} $ and $ s \in \mc V $ such that $|N_s^-(x)| \geq 1$. Let $k \in N_s^-(x)$ and set $ T_1 = ( 1-\frac{\beta}{2}) (1-\beta)^{-1}(\eta_s + \frac{\beta}{2}\eta_{k})^{-1} $ and $ T_2 = (1-\beta)^{-1}(\eta_s + \frac{\beta}{2} \eta_k)^{-1} $. Then it holds that:
\begin{enumerate}
\item $ \tau_{k}^s(x) \leq T_1$ and for all $i \neq k$, $ \tau_{i}^s(x)\leq T_2 $;
\item if $\tau_{k}^s(x) = T_1$ and for all $ i\neq k,s$, $\tau_{i}^s(x) = T_2 $, then $|N^{-2}_s(x)|= 1$.
\end{enumerate}
\end{lem}

\begin{pf}
$ (1) $ 
Let $ \tau_{\text{max}}^s= \max_{j \in \mc V} \;\tau_j^s$. 
By system (\ref{eq:system_tau}) it holds that $\tau_{\text{max}}^s \leq 1  + (1-\beta)\eta_{k}\tau_{k}^s +(1-\beta)(1-\eta_s - \eta_{k})\tau_{\text{max}}^s + \beta \tau_{\text{max}}^s$,
which implies that
\begin{equation}
\label{eq:lemmaUpperBound2_eq2}
\tau_{\text{max}}^s \leq (1-\beta)^{-1}(\eta_{k} + \eta_s)^{-1} + \eta_{k}(\eta_{k}+\eta_s)^{-1}\tau_k^s.
\end{equation}
At the same time, by system (\ref{eq:system_tau}) it holds that $ \tau_{k}^s \leq 1 + (1-\beta)(1-\eta_s - \eta_{k}) \tau_{\text{max}}^s + (1-\beta)\eta_{k}\tau_{k}^s + (\beta/2) \tau_{\text{max}}^s$,
which implies that
\begin{equation}
\label{eq:lemmaUpperBound2_eq3}
\tau_{k}^s \leq \frac{1+\left[(1-\beta)(1-\eta_{k}-\eta_s)+\frac {\beta}{2}\right]\tau_{\text{max}}^s}{1 -(1-\beta)\eta_{k}}.
\end{equation}
By substituting inequality (\ref{eq:lemmaUpperBound2_eq2}) in (\ref{eq:lemmaUpperBound2_eq3}), the following upper bound is obtained:
\[
\tau_{k}^s\leq T_1 = \left(1-\frac{\beta}{2}\right)(1-\beta)^{-1}\left(\eta_s + \frac{\beta}{2}\eta_{k}\right)^{-1},
\]
while by substituting inequality  (\ref{eq:lemmaUpperBound2_eq3}) in (\ref{eq:lemmaUpperBound2_eq2}) we obtain:
\[
\tau_{\text{max}}^s\leq T_2 = (1-\beta)^{-1}\left(\eta_s + \frac{\beta}{2} \eta_k\right)^{-1}.
\]

(2) Suppose that there exists $ j\neq k $ such that $j\in N_s^-(x) \cup N_k^-(x) $; we show that this leads to a contradiction.  
There are three cases: either $ x_j=\lbrace s,k \rbrace $, or there exists $ b\neq k,s $ such that $ x_j=\lbrace s,b\rbrace $ or $ x_j=\lbrace k,b\rbrace $. By system (\ref{eq:system_tau}), $\tau_j^s$ satisfies:
\begin{equation}
    \label{eq:Lemma11_item2_1}
    \tau_j^s \leq 1 + (1-\beta)\sum_{i \in \mc V}\eta_i \tau_i^s + \frac{\beta}{2}\left(\tau_{k}^s + \tau_b^s\right).
\end{equation}
By substituting the values of the $ \tau_{i}^s $'s in the hypothesis and by observing that $ T_1<T_2 $, equation (\ref{eq:Lemma11_item2_1}) leads to:
\begin{align*}
    T_2 &\leq 1 + (1-\beta)\left( T_2 + \eta_k T_1 -(\eta_k + \eta_s)T_2\right) + \frac{\beta}{2}\left(T_1 + T_2\right) \\
    &< 1 + (1-\beta)\left( T_2 + \eta_k T_2 -(\eta_k + \eta_s)T_2\right) + \frac{\beta}{2}\left(2 T_2\right) \\
    &< \left[ (1-\beta) \frac{\beta}{2} \eta_k + 1\right] (1-\beta)^{-1}\left(\eta_s + \frac{\beta}{2} \eta_k\right)^{-1} <  T_2,
\end{align*}
which is a contradiction. This means that the set $N_s^-(x) \cup N_k^-(x)$ has to be equal to $\{k\}$ and so $|N_s^{-2}(x)|= 1$. $ \qed $
\end{pf}

The following proposition characterizes the best response set of a player in the game $ \Gamma(\mc V,\beta,\eta, 2) $ and it will play a key role in both the proofs of Theorem \ref{thm:strictNash_m2} and \ref{thm:condensation_graph}.
From now on, fixed $ s\in\mc V $ and $ x\in \mc A $, we label the elements of $ \mc V $ in such a way that $ \mc V=\lbrace s, v_1,\dots ,v_{n-1}\rbrace $ and 
\begin{equation}\label{eq:tau_v}
0=\tau_s^s(x)<\tau_{v_1}^s(x)\leq \tau_{v_2}^s(x)\leq \dots \leq \tau_{v_{n-1}}^s(x).
\end{equation}


\begin{prop}\label{prop:bestresponse_m2}
Consider the game $ \Gamma(\mc V,\beta,\eta, 2) $, and let $ x\in\mc A $ and $ s\in \mc V $. It holds that:
\begin{enumerate}
\item if $ N_s^{-2}(x)\!=\! \emptyset $, then $ \mathcal{B}_s(x_{-s})\!=\!\bigl\lbrace \lbrace v,w\rbrace\! : v,w\!\in\! \mc V \setminus\lbrace s\rbrace, v\neq w \bigr\rbrace $;
\item if $ |N_s^{-2}(x)|\!=\!1 $, then $ \mathcal{B}_s(x_{-s})\!=\!\bigl\lbrace \lbrace r,v\rbrace \!: v\in \mc V \setminus\lbrace s,r\rbrace \bigr\rbrace $, where $ \lbrace r\rbrace=N_s^{-2}(x)=N_s^{-}(x) $;
\item if $ |N_s^{-2}(x)|\!\geq\! 2 $, then $ \mathcal{B}_s(x_{-s})\subseteq
\bigl\lbrace \lbrace v,w\rbrace \!: v,w\in N^{-}_s(x), v\!\neq\! w\bigr\rbrace \cup  \bigl\lbrace \lbrace v,w\rbrace \!: v\!\in\! N^{-}_s(x) \text{ and } w\!\in\! N^{-}_v(x)\bigr\rbrace$.
\end{enumerate}
\end{prop}

\begin{pf}
$ (1) $ If $N_s^{-2}(x)= \emptyset $, then $ \tau_j^s $ can still be expressed as in (\ref{eq:tau_emptyinneigh}), so we conclude.\\
$ (2) $ We remind that we label the elements of $ \mc V $ in such a way that (\ref{eq:tau_v}) holds.
We first show that $v_1\in N_s^{-}(x) $. By contradiction, suppose that $v_1\notin N_s^{-}(x) $; then $ x_{v_1}=\lbrace a,b\rbrace $ for some $ a,b \neq s $. It holds that
\[
\tau_{v_1}^s=1+(1-\beta)\sum_{v\in \mc V}\eta_v\tau_v^s+\frac{\beta}{2}(\tau_a^s+\tau_b^s)\geq 1+\tau_{v_1}^s-\eta_s(1-\beta)\tau_{v_1}^s, 
\]
which implies that $ \tau_{v_1}^s\geq \eta_s^{-1}(1-\beta)^{-1} $. By Lemma \ref{lem:upperbound_tau}, it follows that $ \tau_{v_1}^s=\eta_s^{-1}(1-\beta)^{-1} $ and $N_s^{-}(x)= \emptyset $, which is a contradiction. Therefore, if $ N_s^{-2}(x)=N_s^{-}(x)=\lbrace r\rbrace $, it holds that $ r=v_1 $ and so $ r\in x_s $ for any $ x_s\in \mc B_s(x_{-s}) $. We now show that $ \tau_j^s=\tau_k^s $ for every $ j,k\neq r,s $, which implies that $ \mathcal{B}_s(x_{-s})=\bigl\lbrace \lbrace r,v\rbrace : v\in \mc V \setminus\lbrace s,r\rbrace \bigr\rbrace $. 
By hypothesis, for every $j\neq s,r $, the probability to jump from $ j $ to $ s $ is equal to $ (1-\beta)\eta_s $ and the probability to jump from $ j $ to $ r $ is equal to $ (1-\beta)\eta_r$. It follows that the probability to arrive in $ s $ from $ j $ in exactly $ t $ steps without passing through $ r $ is equal to $(1-\beta)\eta_s(1-(1-\beta)(\eta_s+\eta_r))^{t-1}  $ and the probability to arrive in $ r $ from $ j $ in exactly $ t $ steps without passing through $ s $ is equal to $(1-\beta)\eta_r(1-(1-\beta)(\eta_s+\eta_r))^{t-1}  $. Consequently,
\[
\tau_j^s=\sum_{t=1}^{\infty}(1-\beta)(t\eta_s+\eta_r(t+\tau_r^s))\bigl( 1-(1-\beta)(\eta_s+\eta_r)\bigr)^{t-1}, 
\]
which does not depend on $ j $.\\ 
$ (3) $ Suppose that $ |N_s^{-2}(x)|\geq 2 $. We already proved that $ v_1\in N^{-}_s(x) $; we need to prove that either $ v_2\in  N^{-}_s(x)$ or $ v_2\in  N^{-}_{v_1}(x) $. Suppose by contradiction that this is not the case and let $ a,b\neq s,v_1 $ such that $ x_{v_2}=\lbrace a,b\rbrace $. By applying system (\ref{eq:system_tau}) to express $ \tau_{v_2}^s $ and by using the fact that for all $ j\geq 2$, $ \tau_{v_j}^s\geq \tau_{v_2}^s $, it holds that:
\begin{equation}
\label{eq:PropUpperBound2_eq1}
\tau_{v_2}^s \geq \frac{1}{(1-\beta)(\eta_{v_1} + \eta_s)} + \frac{\eta_{v_1}}{\eta_{v_1}+\eta_s}\tau_{v_1}^s.
\end{equation}
Moreover, by applying system (\ref{eq:system_tau}) to express $ \tau_{v_1}^s $ and by using again the fact that for all $ j\geq 2$, $ \tau_{v_j}^s\geq \tau_{v_2}^s $, it holds that:
\begin{equation}
\label{eq:PropUpperBound2_eq2}
\tau_{v_1}^s \geq \frac{1+\left[(1-\beta)(1-\eta_{v_1}-\eta_s)+\frac {\beta}{2}\right]\tau_{v_2}^s}{1 -(1-\beta)\eta_{v_1}}.
\end{equation}
By substituting inequality (\ref{eq:PropUpperBound2_eq1}) in (\ref{eq:PropUpperBound2_eq2}) and inequality (\ref{eq:PropUpperBound2_eq2}) in (\ref{eq:PropUpperBound2_eq1}) we obtain respectively:
\[
\tau_{v_1}^s\geq T_1 \quad \text{and}\quad \tau_{v_2}^s\geq T_2,
\]
where $ T_1 $ and $ T_2 $ are defined in Lemma \ref{lem:upperbound2_tau}. Therefore, by (\ref{eq:tau_v}) and  item (1) of Lemma \ref{lem:upperbound2_tau}, it holds that $\tau_{v_1}^s = T_1$, and for all $ j \geq 2 $, $\tau_{v_j}^s = T_2$. By applying item (2) of the same lemma it follows that $|N_s^{-2}(x)| = 1$, which contradicts the hypothesis. $\qquad\qquad\qquad\qquad\qquad\qquad\qquad\qquad\quad \qed $
\end{pf}


\begin{rem}
\label{rem:viciniAdist1-viciniAdist2}
Suppose that $|N_s^{-2}(x)|\!\geq\! 2 $ for some $ x\in\mc A $ and $ s\in \mc V $ and let $ x_s=\lbrace i,j\rbrace\in  \mathcal{B}_s(x_{-s})$.
Item (3) of Proposition \ref{prop:bestresponse_m2} implies that, if $ j\notin N^{-}_s(x) $, then $ i\in N^{-}_s(x)  $ and $j\in N^{-}_i(x)$. In other words, if
$ j $ is not an in-neighbor of $ s $, then $ (j,i) $ and $ (i,s) $ must be edges of $ \mc G(x) $, together with the edges $ (s,i) $ and $ (s,j) $ as $ s $ is playing $ \lbrace i,j\rbrace $.
\end{rem}

\begin{figure}
\begin{center}
\begin{tikzpicture}[roundnode/.style={circle,fill=black!30, inner sep=1pt, minimum size=3mm},node distance=1cm,line width=0.2mm,  ]

\node[roundnode] (q_1) at (0,0){$s$};
\node[roundnode]  (q_2) at (1,0){};
\node[roundnode] (q_3) at (2,0){};
\node[roundnode] (q_4) at (1.5,-1){};
\node[roundnode] (q_5) at (1.5,1){};

\node[roundnode] (q_6) at (2,2){};
\node[roundnode] (q_7) at (1,2){};

\path[-]
(q_7)  edge []		node  {} (q_6)
(q_4) edge []		        node  {} (q_5) 
; 

\path[->] 
 (q_1) edge []		        node  {} (q_2)
 (q_1) edge []		        node  {} (q_4)
 (q_7)  edge []		node  {} (q_5)
 (q_6)  edge []             	    	node  {} (q_5)
 (q_2)  edge []		    node  {} (q_4)
 (q_2)  edge []		    node  {} (q_5)
 (q_3) edge []		        node  {} (q_2)
 (q_3) edge []		        node  {} (q_5)
 (q_4) edge []		        node  {} (q_3)         		  
 ; 
  
\draw [dashed] (0.7,-1.3) rectangle (2.3,2.3);  
\node at (1.5,-1.7) {$ \mc B_s(x_{-s}) $};
\node at (0,-1.3) {(a)}; 
  
\end{tikzpicture}
\begin{tikzpicture}[roundnode/.style={circle,fill=black!30, inner sep=1pt, minimum size=3mm},node distance=1cm,line width=0.2mm,  ]

\node[roundnode] (q_1) at (0,0){$s$};
\node[roundnode]  (q_2) at (1,0){};
\node[roundnode] (q_3) at (2,0){};
\node[roundnode] (q_4) at (1.5,-1){};
\node[roundnode] (q_5) at (1.5,1){};

\node[roundnode] (q_6) at (2,2){};
\node[roundnode] (q_7) at (1,2){$ r $};

\path[-]
(q_2)  edge []		    node  {} (q_4)
 (q_6)  edge []             	    	node  {} (q_3)
 (q_2)  edge []		    node  {} (q_3)
 (q_1) edge []		        node  {} (q_7)
;
\path[->] 
 (q_1) edge []		        node  {} (q_4)
 (q_7)  edge []		node  {} (q_5)
 (q_6)  edge []             	    	node  {} (q_5)
 (q_5)  edge []	    	node  {} (q_2)
 (q_5)  edge []	    	node  {} (q_3)
 (q_4) edge []		        node  {} (q_3)
           		  
 ; 
  
\draw [dashed] (0.7,1.7) rectangle (1.3,2.3);  
\draw [dashed] (0.6,-1.3)--(0.6,1.5)--(1.5,1.5)--(1.5,2.4)--(2.3,2.4)--(2.3,-1.3)--(0.6,-1.3);  
\node at (1.5,-1.7) {$ \mc B_s(x_{-s}) $};
\node at (0,-1.3) {(b)};  
\end{tikzpicture}
\begin{tikzpicture}[roundnode/.style={circle,fill=black!30, inner sep=1pt, minimum size=3mm},node distance=1cm,line width=0.2mm,  ]

\node[roundnode] (q_1) at (0,0){$s$};
\node[roundnode]  (q_2) at (1,0){};
\node[roundnode] (q_3) at (2.3,0){};
\node[roundnode] (q_4) at (1.5,-1){};
\node[roundnode] (q_5) at (1.5,1){};

\node[roundnode] (q_6) at (2.3,2){};
\node[roundnode] (q_7) at (1,2){};

\path[-]
(q_1) edge []		        node  {} (q_2)
(q_5)  edge []	    	node  {} (q_3)
(q_3) edge []		        node  {} (q_6)
 (q_5)  edge []	    	node  {} (q_7)
;

\path[->] 
 (q_1) edge []		        node  {} (q_4)
 (q_7)  edge []		node  {} (q_1)
 (q_6)  edge []             	    	node  {} (q_5)
 (q_2)  edge []		    node  {} (q_3)
 (q_4) edge []		        node  {} (q_3)
 (q_4) edge []		        node  {} (q_2)          		  
 ; 
  
\draw [dashed] (0.6,-1.3) rectangle (1.8,2.2); 

\node at (1,-1.7) {$ \mc B_s(x_{-s}) \subseteq $}; 
\node at (-0.3,-1.3) {(c)}; 
\end{tikzpicture}
\end{center}
\caption{The best response set $ \mc B_s(x_{-s}) $ of player $ s $ in the game $ \Gamma(\mc V,\beta,\eta, 2) $ when $ N^-_s(x) $ is (a) empty (b) of cardinality one (c) of cardinality greater than one. In case (b), node $s$ has to link to its unique in-neighbor $r$ and then it can place its second link anywhere else.} 
\label{fig:bestresponse_m2}
\end{figure}
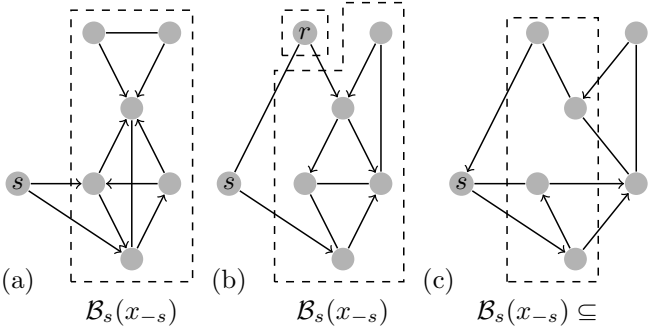

Figure \ref{fig:bestresponse_m2} graphically synthesizes Proposition \ref{prop:bestresponse_m2}.
Notice that in view of Proposition \ref{prop:bestresponse_m2}, the best response of a node $ s $ can be unique only in the case $ |N_s^{-2}(x)|\geq 2 $.

\begin{figure}
\begin{center}
(a)
\begin{tikzpicture}[roundnode/.style={circle,fill=black!30, inner sep=0.5pt, minimum size=6mm},node distance=1cm,line width=0.2mm, ]

\node[regular polygon, regular polygon sides=6,minimum size=2.2cm] (esa) at (0,0) {};
\node[roundnode] (q_1) at (esa.corner 1) {2};
\node[roundnode] (q_2) at (esa.corner 2) {1};
\node[roundnode] (q_3) at (esa.corner 3) {s};
\node[roundnode] (q_4) at (esa.corner 4) {n-1};
\node[roundnode] (q_5) at (esa.corner 5) {n-2};
\node[roundnode] (q_6) at (esa.corner 6) {$ \dots $};

\path[-] (q_1) edge 		node  {} (q_2)
 (q_2) edge 	node  {} (q_1)
 (q_2) edge 	node  {} (q_3)
 (q_3) edge 		node  {} (q_2)
 (q_3) edge 		node  {} (q_4)
 (q_4) edge 	node  {} (q_3)
 (q_4) edge 	node  {} (q_5)
 (q_5) edge 		node  {} (q_4)
 (q_5) edge [dotted]		node  {} (q_6)
 (q_1) edge [dotted]		node  {} (q_6)
 ; 
\end{tikzpicture}
$ \qquad $(b)
\begin{tikzpicture}[roundnode/.style={circle,fill=black!30, inner sep=2pt, minimum size=6mm},node distance=0.7cm,line width=0.2mm]
 			\node[roundnode]    (q_1) at (1.5,0) {j};
 			\node[roundnode]    (q_2)  at (0,0) {s};
 			\node[roundnode]          (q_3) at (0,1.5) {i};			
 			
 \path[->] (q_2) edge 		node  {} (q_1);
 \path[-] (q_1) edge		node  {} (q_3)
 (q_3) edge		node  {} (q_2)
 ;
  
\end{tikzpicture}
\end{center}

\caption{(a) A ring graph on $ n $ nodes. (b) The directed graph $T_{(s,j),i}$. }
\label{fig:cycle_graph}
\end{figure}
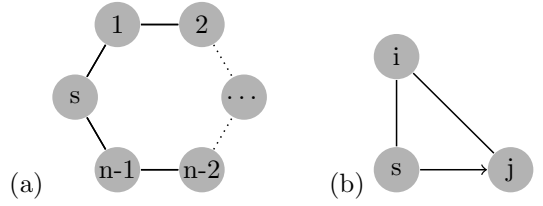

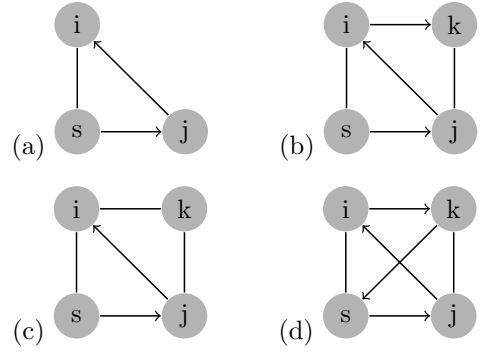
\begin{figure}
\begin{center}
(a)
\begin{tikzpicture}[roundnode/.style={circle,fill=black!30, inner sep=1pt, minimum size=6mm},node distance=0.8cm,line width=0.2mm]

\node[roundnode] (q_1) {i};
\node[roundnode] (q_2) [below=of q_1] {s};
\node[roundnode] (q_3) [right=of q_2] {j};

\path[-] (q_1) edge 		node  {} (q_2)
 ;

\path[->] 
 (q_2) edge		node  {} (q_3)
 (q_3) edge 		node  {} (q_1)
 ; 
\end{tikzpicture}
$ \qquad $
(b)
\begin{tikzpicture}[roundnode/.style={circle,fill=black!30, inner sep=1pt, minimum size=6mm},node distance=0.8cm,line width=0.2mm]

\node[roundnode] (q_1) {i};
\node[roundnode] (q_2) [below=of q_1] {s};
\node[roundnode] (q_3) [right=of q_2] {j};
\node[roundnode] (q_4) [right=of q_1] {k};

\path[-] (q_1) edge 		node  {} (q_2)
  (q_3) edge		node  {} (q_4)
 ; 

\path[->] 
 (q_2) edge		node  {} (q_3)
 (q_3) edge 		node  {} (q_1)
 (q_1) edge 		node  {} (q_4)
 ; 
\end{tikzpicture}\\
$  $\\
(c)
\begin{tikzpicture}[roundnode/.style={circle,fill=black!30, inner sep=1pt, minimum size=6mm},node distance=0.8cm,line width=0.2mm]

\node[roundnode] (q_1) {i};
\node[roundnode] (q_2) [below=of q_1] {s};
\node[roundnode] (q_3) [right=of q_2] {j};
\node[roundnode] (q_4) [right=of q_1] {k};

\path[-] (q_1) edge 		node  {} (q_2)
  (q_3) edge		node  {} (q_4)
 (q_4) edge 	node  {} (q_1)
 ; 

\path[->] 
 (q_2) edge		node  {} (q_3)
 (q_3) edge 		node  {} (q_1)
 ; 
\end{tikzpicture}
$ \qquad $
(d)
\begin{tikzpicture}[roundnode/.style={circle,fill=black!30, inner sep=1pt, minimum size=6mm},node distance=0.8cm,line width=0.2mm]

\node[roundnode] (q_1) {i};
\node[roundnode] (q_2) [below=of q_1] {s};
\node[roundnode] (q_3) [right=of q_2] {j};
\node[roundnode] (q_4) [right=of q_1] {k};

\path[-] (q_1) edge		node  {} (q_2)
  (q_3) edge 		node  {} (q_4)
 ;

\path[->] 
 (q_2) edge		node  {} (q_3)
 (q_3) edge 		node  {} (q_1)
 (q_4) edge	node  {} (q_2)
 (q_1) edge 		node  {} (q_4)
 ; 
\end{tikzpicture}
\end{center}
\caption{Explanatory graphs for the proof of Theorem \ref{thm:strictNash_m2}.}
\label{fig:proof}
\end{figure}

\textbf{Proof of Theorem \ref{thm:strictNash_m2}.} 
We first prove that a ring graph on $ n $ vertices is a strict Nash equilibrium for $ \Gamma(\mc V,\beta, \eta,2) $. If $ n=3 $ the proof is trivial. Suppose that $ n>3  $ and 
consider the ring graph as in Fig.\ \ref{fig:cycle_graph}(a); we want to show that the node $ s $ is in its (unique) best response, that is we want to show that $ \tau_1^s, \tau_{n-1}^s< \tau_v^s $ for all $ v\neq 1, n-1 $. By the symmetry of the graph, $ \tau_1^s=\tau_{n-1}^s$ and $ \tau_2^s=\tau_{n-2}^s$. In view of Remark \ref{rem:viciniAdist1-viciniAdist2}, it then suffices to show that $ \tau_1^s<\tau_2^s $. By system (\ref{eq:system_tau}), we have that $\tau_{2}^s-\tau_{1}^s= (\beta/2) (\tau_1^s-\tau_2^s)+(\beta/2)\tau_3^s$, which implies that $ \tau_2^s>\tau_1^s $ since $ \tau_3^s>0 $.	\\
We now show that if $ x^* $ is a strict Nash equilibrium for $ \Gamma(\mc V,\beta, \eta,2) $, then $\mathcal{G}(x^*)$ is undirected, which implies that $\mathcal{G}(x^*)$ is the union of ring graphs since by construction each vertex of $\mathcal{G}(x^*)$ has out-degree equal to $ 2 $.
Assume by contradiction that there exists a strict Nash equilibrium $ x\in \mc A $ and two nodes $ s,j\in \mc V $ such that $ (s,j)\in \mc E(x)$ but $  (j,s)\notin \mc E(x) $. Since $ x $ is a strict Nash equilibrium, all the nodes are in their best response and $ |\mathcal{B}_v(x)|=1 $ for all $ v\in \mc V $. By Proposition \ref{prop:bestresponse_m2} we know that $  j\in N^{-2}_s(x) $: since $(j,s)\notin \mc E(x) $, it means that there exists $ i\neq j,s $ such that $(j,i),(i,s),(s,i)\in \mc E(x)$ by Remark \ref{rem:viciniAdist1-viciniAdist2} (see also Fig.\ \ref{fig:proof}(a)).  This also implies that $  i\in N^{-2}_j(x) $. 
If $ i\in N^{-}_j(x) $, by system (\ref{eq:system_tau}) it holds that $ \tau_i^j-\tau_s^j=(\beta/2)(\tau_s^j-\tau_i^j)$ and so $\tau_i^j=\tau_s^j  $. Therefore we have that either $ (j,s)\in \mc E(x) $ or $ |\mathcal{B}_j(x)|>1 $, both cases leading to a contradiction. We now examine the case $ i\in N^{-2}_j(x)\setminus N^{-}_j(x) $: by Remark \ref{rem:viciniAdist1-viciniAdist2} there exists $ k\neq i,j $ such that $(i,k),(k,j),(j,k)\in \mc E(x) $ (see Fig.\ \ref{fig:proof}(b)). Proposition \ref{prop:bestresponse_m2} also implies that $ k\in N^{-2}_i(x) $. If $ k\in N^{-}_i(x) $, we are in the situation represented in Fig.\ \ref{fig:proof}(c); by using system (\ref{eq:system_tau}), it is easy to see that $ \tau_s^j=\tau_k^j $. This implies that either $ k=s $ (in which case $ (j,s)\in \mc E(x) $) or $ |\mathcal{B}_j(x)|>1 $, so we always arrive to a contradiction. Finally, we need to consider the case $ k\in N^{-2}_i(x)\setminus N^{-}_i(x) $: since the actions of $ i $ are determined as in Fig.\ \ref{fig:proof}(b), it must hold that $ (k,s)\in \mc E(x) $, as represented in Fig.\ \ref{fig:proof}(d). By using again system (\ref{eq:system_tau}) to express $ \tau_i^j $ and $ \tau_s^j $, we get that $(1+\beta/2)( \tau_i^j -\tau_s^j)=(\beta/2)\tau_k^j>0 $ and so $ \tau_i^j >\tau_s^j $. This implies that $ j $ is not in its best response, thus leading to a contradiction. $ \qquad\qquad\qquad\qquad\qquad\qquad\qquad\qquad\qed $

Before proving Theorem \ref{thm:condensation_graph}, we first need the following definition and Lemma \ref{lem:triangle}.

\begin{defn}\label{defn:triangle}
We denote by $T_{(s,j),i}$ the directed graph on the vertices $ \lbrace i,j,s\rbrace $ having one directed edge $ (s,j) $ and all the other edges undirected (see Fig.\ \ref{fig:cycle_graph}(b)). We will sometimes refer to $T_{(s,j),i}$ as a \emph{triangle}.
\end{defn}

\begin{lem}\label{lem:triangle}
Let $ x\in \mc A $ be a Nash equilibrium for the game $ \Gamma(\mc V,\beta, \eta,2) $, $\mc H(x)$ be the condensation graph of $ \mc G(x) $ and let $\mc G_\lambda(x)=(\mc V_\lambda,\mc E_\lambda)$ be a sink in $\mc H(x)$. If there exists $(s,j)\in\mc E_{\lambda}$ that is directed, then $\mc G_\lambda(x)$ contains a structure of type $T_{(s,j),i}$.
\end{lem}

\begin{pf} Notice that since the out-degree of each node in $\mc G_\lambda(x)$ is equal to two, this graph must contain at least three nodes and $ |N_s^{-2}(x)|\geq 2$. It follows that $j\in N_s^{-2}(x)$ and so by Remark \ref{rem:viciniAdist1-viciniAdist2}, there exists $i\in\mc V_{\lambda}$ such that $(j,i), (i,s), (s,i)\in\mc E_{\lambda}$ (see Fig.\ \ref{fig:proof}(a)). We are left to prove that $(i,j)\in\mc E_{\lambda}$. If this was not the case, then by Remark \ref{rem:viciniAdist1-viciniAdist2} there would exist $ k\in\mc V_\lambda $ such that $(i,k), (k,j), (j,k)\in\mc E_{\lambda}$, i.e.\ the graph in Fig.\ \ref{fig:proof}(b) would be a subgraph of $\mc G_\lambda(x)$.
In this configuration, the only way $ i$ could be at equilibrium is that $ (k,i)\in \mc E_\lambda $, as otherwise $\lbrace s,j\rbrace$ would give it a strictly better utility than $\lbrace s,k\rbrace$. We would then be in the configuration of Fig.\ \ref{fig:proof}(c); but in this case $ j $ is not at equilibrium, as $\lbrace s,k\rbrace$ gives it a strictly better utility than $\lbrace i,k\rbrace$. This completes the proof. $\qquad\qquad\qquad\qquad \qed $
\end{pf}

We are now ready to prove Theorem \ref{thm:condensation_graph}.

\begin{figure}
\begin{center}
\begin{tikzpicture}[roundnode/.style={circle,fill=black!30, inner sep=2pt, minimum size=5mm},node distance=0.7cm,line width=0.2mm]
 			\node[roundnode]    (q_1) at (0,0) {s};
 			\node[roundnode]    (q_2)  at (0,-2) {i};
 			\node[roundnode,fill=black]          (q_3) at (1.5,-1) {\textcolor{white}{j}};
 			\node[roundnode]          (q_4) at (3,-1) {};	
 			\node[roundnode]          (q_5) at (5,-1) {};
 			\node[roundnode,fill=black]          (q_6) at (6.5,-1) {\textcolor{white}{k}};
 			\node[roundnode]          (q_7) at (8,0) {r};
 			\node[roundnode]          (q_8) at (8,-2) {t};			
 			
 \path[->] (q_7) edge 		node  {} (q_6) 
(q_1) edge		node  {} (q_3)
 ;
 \path[-] 
 (q_3) edge		node  {} (q_2)
 (q_1) edge 		node  {} (q_2)
 (q_4) edge		node  {} (q_3)
 (q_6) edge		node  {} (q_5)
 (q_6) edge		node  {} (q_8)
 (q_8) edge		node  {} (q_7)
 ;
  \path[dashed] (q_4) edge		node  {} (q_5);
\end{tikzpicture}
\end{center}
\caption{Explanatory graph for the proof of Theorem \ref{thm:condensation_graph}. Black nodes are not in best response.}
\label{fig:long}
\end{figure}
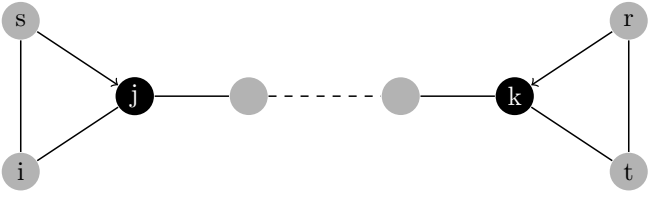

\textbf{Proof of Theorem \ref{thm:condensation_graph}.}
Consider any component $\mc G_\lambda(x)=(\mc V_\lambda,\mc E_\lambda)$ that is not a sink in $\mc H(x)$. Necessarily, there must exist $i\in \mc V_\lambda$ such that $N_i(x)\not\subseteq \mc V_{\lambda}$. In particular, this implies that 
$ |N_i^{-2}(x)|\leq 1$ by Proposition \ref{prop:bestresponse_m2}. If $ |N_i^{-2}(x)|=0$, it means that $\mc V_\lambda=\{i\}$ is a singleton. If $ |N_i^{-2}(x)|=1$, then necessarily $\mc V_\lambda=\{i,j\}$ for some $j\neq i$ and so $\mc G_\lambda(x)$ is the $2$-clique on $\{i,j\}$. Notice that in both cases, there cannot be any other component $\mc G_{\lambda'}(x)$ linking to $\mc G_{\lambda}(x)$ in the condensation graph, as otherwise the condition $ |N_i^{-2}(x)|\leq 1$ would be violated. This proves items $(1)$ and $(2)$. \\
We now study the structure of the sink components. Suppose that the component $\mc G_\lambda(x)=(\mc V_\lambda,\mc E_\lambda)$ is not a ring graph and thus not undirected; then there must exist at least two directed edges in $\mc E_{\lambda}$. Let $(s,j)$ be one of these directed edges and let $T_{(s,j),i}$ be the corresponding triangle (see Definition \ref{defn:triangle} and Lemma \ref{lem:triangle}). We now discuss how any other triangle $T_{(r,k), t}$ in $\mc G_\lambda(x)$ can possibly intersect with $T_{(s,j),i}$. Notice that, since the out-degree of all nodes in $\mc G_\lambda(x)$ is $2$, the two triangles cannot intersect in the nodes of out-degree equal to two in the corresponding triangles, namely 
$\{i,s\}\cap\{r,k,t\}=\emptyset$ and $\{r,t\}\cap\{i,j,s\}=\emptyset$. Therefore the only possibility is that they have just one node in common, namely $j=k$; this corresponds to the Butterfly graph (see Fig. \ref{fig:C2_nash}(b)). Since in the Butterfly graph every node has out-degree equal to $2$, it necessarily coincides with the connected component $\mc G_\lambda(x)$. If instead $T_{(s,j),i}$ does not intersect any other triangle, there must exist a sequence of distinct nodes $j=j_1, j_2,\dots , j_l=r$, with $l\geq 2$, such that $\{j_a,j_{a+1}\}$ are $2$-cliques in $\mc G_\lambda(x)$ for $a=1,\dots ,l-1$ and such that there exists a triangle $T_{(r,k), t}$ in $\mc G_\lambda(x)$ for some $ k,t $. Since there cannot be any incoming directed edge in $r$ by hypothesis, we deduce that $N_r^{-2}(x)=\{j_{l-1}, j_{l-2}\}$ if $l\geq 3$ and $N_r^{-2}(x)=\{i,j,s\}$ if $l=2$. This last case is impossible since it would result that $k\in \{i,j,s\}$, contrarily to what we had assumed. In the case when $l\geq 3$, we obtain that $k=j_{l-2}$ that leads to the graph depicted in Fig. \ref{fig:long}. 
A direct computation shows that nodes $j$ and $k$ are however not at equilibrium in this configuration. This completes the proof. $\qquad\qquad\qquad\qquad\qquad\qquad\quad \qed $

We are left to prove Remark \ref{remark_m2} and Corollary \ref{cor:trapping_setsM2}.

\textbf{Proof of Remark \ref{remark_m2}.}
(1) Consider Fig.\ \ref{fig:singleton_ring}(a); we show that node $ s $ is not playing an action in its best response set. Indeed by system (\ref{eq:system_tau}), it holds that $\tau_j^{s}- \tau_{1}^{s} = \frac{\beta}{2} (\tau_{1}^{s}- \tau_{2}^{s})$. In the proof of Theorem \ref{thm:strictNash_m2} we showed that if the ring has length greater or equal than four, then $\tau_1^{s} < \tau_{2}^{s}$ and therefore $\tau_j^{s}< \tau_{1}^{s}$. It follows that node $s$ is not playing a best response action and so such configuration is not a Nash equilibrium.\\
(2) Consider Fig.\ \ref{fig:singleton_ring}(b); we show that node $ s $ is not playing an action in its best response set. 
By symmetry $\tau_j^s = \tau_{k}^{s}$, so it sufficies to show that $ \tau_1^s< \tau_j^s $. By system (\ref{eq:system_tau}), $(1+\beta/2)(\tau_{1}^{s}- \tau_{j}^{s}) = (\beta/2) (\tau_{2}^{s}- \tau_{1}^{s})$; in the proof of Theorem \ref{thm:strictNash_m2} we showed that $\tau_{2}^{s} > \tau_{1}^{s}$, so we conclude.$\qquad\qquad \qed $

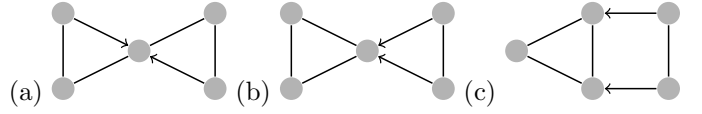
\begin{figure}
\centering     
(a)
\begin{tikzpicture}[roundnode/.style={circle,fill=black!30, inner sep=2pt, minimum size=3mm},node distance=0.3cm,line width=0.2mm,scale=0.5]
 			\node[roundnode]    (q_1) at (0,0) {};
 			\node[roundnode]    (q_2)  at (0,-2) {};
 			\node[roundnode]          (q_3) at (2,-1) {};
 			\node[roundnode]          (q_4) at (4,0) {};	
 			\node[roundnode]          (q_5) at (4,-2) {};			
 			
 \path[->] (q_5) edge 		node  {} (q_3) 
(q_1) edge		node  {} (q_3)
 ;
 \path[-] 
 (q_3) edge		node  {} (q_2)
 (q_1) edge 		node  {} (q_2)
 (q_4) edge		node  {} (q_3)
 (q_4) edge		node  {} (q_5)
 ;
  
\end{tikzpicture}
(b)
\begin{tikzpicture}[roundnode/.style={circle,fill=black!30, inner sep=2pt, minimum size=3mm},node distance=0.3cm,line width=0.2mm,scale=0.5]
 			\node[roundnode]    (q_1) at (0,0) {};
 			\node[roundnode]    (q_2)  at (0,-2) {};
 			\node[roundnode]          (q_3) at (2,-1) {};
 			\node[roundnode]          (q_4) at (4,0) {};	
 			\node[roundnode]          (q_5) at (4,-2) {};			
 			
 \path[->] (q_5) edge 		node  {} (q_3) 
(q_4) edge		node  {} (q_3)
 ;
 \path[-] 
 (q_3) edge		node  {} (q_2)
 (q_1) edge 		node  {} (q_2)
 (q_1) edge		node  {} (q_3)
 (q_4) edge		node  {} (q_5)
 ;
  
\end{tikzpicture}
(c)
\begin{tikzpicture}[roundnode/.style={circle,fill=black!30, inner sep=2pt, minimum size=3mm},node distance=0.3cm,line width=0.2mm,scale=0.5]
 			\node[roundnode]    (q_1) at (0,0) {};
 			\node[roundnode]    (q_2)  at (0,-2) {};
 			\node[roundnode]          (q_3) at (-2,-1) {};
 			\node[roundnode]          (q_4) at (2,0) {};	
 			\node[roundnode]          (q_5) at (2,-2) {};			
 			
 \path[->] (q_5) edge 		node  {} (q_2) 
(q_4) edge		node  {} (q_1)
 ;
 \path[-] 
 (q_1) edge		node  {} (q_3)
 (q_3) edge		node  {} (q_2)
 (q_1) edge 		node  {} (q_2)
 
 (q_4) edge		node  {} (q_5)
 ;
  
\end{tikzpicture}

\caption{Transitions of the Butterfly graph.}
\label{fig:trasitionButterfly}
\end{figure}

\textbf{Proof of Corollary \ref{cor:trapping_setsM2}.}
We know that $\mc N^{st} \subseteq \mc N^*$. 
Let $ x\in \mc N^* \setminus \mc N^{st}$ and let $ \mc G(x) $ be its associated graph; $ \mc G(x) $ must have a directed link. 
The first key observation is that the transition states of the Butterfly graph are the ones shown in Fig.\ \ref{fig:trasitionButterfly}, which are all Nash equilibria. Hence, every time the graph $\mc G(x)$ contains a Butterfly graph, there is a nonzero probability that the best response dynamics will assume the configurations (b) or (c) in Fig.\ \ref{fig:trasitionButterfly}, i.e.\ a configuration with a 2-clique linking to a ring of length three.
The second key observation is that $\mc G(x)$ can have at most one singleton or one 2-clique. In fact, since by Proposition \ref{prop:bestresponse_m2} both singletons and nodes in a 2-clique are always playing a best response action independently on the node they are linking to, there is a nonzero probability that they will direct their links to another singleton or node in a 2-clique, which will not be playing a best response action anymore.
Therefore $\mc G(x)$ has either a singleton, a 2-clique or a Butterfly graph, as the Butterfly graph transforms with nonzero probability into a 2-clique linking to a ring. We are left with the following cases:
\begin{description}
    \item[-] $\mc G(x)$ is a collection of rings and a singleton. It follows from Remark \ref{remark_m2} that $\mc G(x)$ cannot have rings with more than three nodes. If all the rings have length three, there is a nonzero probability that the singleton $ s$ will link to two adjacent nodes $ j$ and $i $ of a ring $ \lbrace i,j,k\rbrace $. In this case it is easy to verify that $ \tau_j^i =\tau_s^i$, and so there is a nonzero probability to end up in a configuration as in Fig.\ \ref{fig:proof}(c), which has been proved not to be a Nash equilibrium. It follows that $\mc G(x)$ cannot contain singletons.
    \item[-] $\mc G(x)$ is a collection of rings and a 2-clique. By Remark \ref{remark_m2}, $\mc G(x)$ cannot have rings with more than three nodes, so all the rings have length three. It follows that the 2-clique can either form configurations (b) or (c) in Fig.\ \ref{fig:trasitionButterfly} or configuration (a) in Fig.\ \ref{fig:C3_nonStrictNash3}, which are all Nash equilibria. Hence $\mc G(x) \in \mc G_b^3$. 
    \item[-] $\mc G(x)$ is a collection of rings and a Butterfly graph. As shown in Fig.\ \ref{fig:trasitionButterfly}, there is a nonzero probability to end up in the previous case, which implies that all the rings have length three. Consequently, $\mc G(x) \in \mc G_b^3$.
\end{description}
We just proved that $\mc N^* \setminus \mc N^{st} \subseteq \mc G_b^3 $. At the same time, every $ x\in  \mc G_b^3 $ is a Nash equilibrium, so $ \mc G_b^3 \subseteq \mc N^* \setminus \mc N^{st}$. Hence $\mc N^*= \mc N^{st} \cup \mc G_b^3$, noticing that $\mc G_b^3$ is not empty if and only if $(n\!\mod 3) =1$.
$ \qquad\qquad\qquad\qquad\qquad\qquad\qed $

\section{Conclusion}\label{conclusions}
In this paper we proposed a game in which every node of a network aims at maximizing its Bonacich centrality by choosing where to direct its out-links, whose number is fixed to be equal to $ m $. We have completely characterized the sets $ \mc N^{\text{st}} $, $ \mc N^* $ and $ \mc N $ of Nash equilibria when $ m=1 $ and the sets $ \mc N^{\text{st}} $ and $ \mc N^* $ when $ m=2 $, case in which we have also provided necessary conditions for a configuration $ x $ to be in $ \mc N $. Our results show that the centrality maximization performed by each node tends to create disconnected and undirected networks, partially due to the locality property of the best response actions. In particular, both for $ m=1 $ and $ m=2 $ all the $ m $-regular undirected networks result to be (strict) Nash equilibria. A natural follow-up of our work would be the analysis of Nash equilibria of the game for a general $ m $, possibly in an heterogeneous setting where $ m $ is different for each node. Preliminary numerical experiments show that this tendency to create disconnected networks show up also for bigger $ m $, and that the problem becomes much more complex. In particular, it seems that the set of Nash equilibria depends also on the parameter $ \beta $ and that not all $ m $-regular undirected networks are Nash equilibria.









\bibliography{bibl_centrality_game}             

\begin{thebibliography}{18}
\providecommand{\natexlab}[1]{#1}
\providecommand{\url}[1]{\texttt{#1}}
\providecommand{\urlprefix}{URL }
\expandafter\ifx\csname urlstyle\endcsname\relax
  \providecommand{\doi}[1]{doi:\discretionary{}{}{}#1}\else
  \providecommand{\doi}{doi:\discretionary{}{}{}\begingroup
  \urlstyle{rm}\Url}\fi

\bibitem[{Acemoglu et~al.(2012)Acemoglu, Carvalho, Ozdaglar, and
  Tahbaz-Salehi}]{Acemoglu12}
Acemoglu, D., Carvalho, V.M., Ozdaglar, A., and Tahbaz-Salehi, A. (2012).
\newblock The network origins of aggergate fluctuations.
\newblock \emph{Econometrica}, 80(5), 1977--2016.

\bibitem[{Avrachenkov and Litvak(2006)}]{Avrachenkov06}
Avrachenkov, K. and Litvak, N. (2006).
\newblock The effect of new links on google pagerank.
\newblock \emph{Stoch. Models}, 22, 2006.

\bibitem[{Ballester and Zenou(2006)}]{Ballester06}
Ballester, C. and Zenou, Y. (2006).
\newblock Who's who in networks. wanted: The key player.
\newblock \emph{Econometrica}, 74, 1403--1417.

\bibitem[{Bonacich(1987)}]{PB:87}
Bonacich, P. (1987).
\newblock Power and centrality: {A} family of measures.
\newblock \emph{American J. of Sociology}, 92(5), 1170--1182.

\bibitem[{Brin and Page(1998)}]{SB-LP:98}
Brin, S. and Page, L. (1998).
\newblock The anatomy of a large-scale hypertextual {W}eb search engine.
\newblock \emph{Computer Networks}, 30, 107--117.

\bibitem[{Cominetti et~al.(2018)Cominetti, Quattropani, and
  Scarsini}]{scarsini}
Cominetti, R., Quattropani, M., and Scarsini, M. (2018).
\newblock The buck-passing game.
\newblock \urlprefix\url{https://arxiv.org/abs/1808.03206}.
\newblock Unpublished.

\bibitem[{Como and Fagnani(2015)}]{Como.Fagnani:2015}
Como, G. and Fagnani, F. (2015).
\newblock Robustness of large-scale stochastic matrices to localized
  perturbations.
\newblock \emph{IEEE Trans. on Network Sci. and Eng.}, 2(2), 53--64.

\bibitem[{Cs{\'a}ji et~al.(2010)Cs{\'a}ji, Jungers, and Blondel}]{Jungers10}
Cs{\'a}ji, B.C., Jungers, R.M., and Blondel, V.D. (2010).
\newblock Pagerank optimization in polynomial time by stochastic shortest path
  reformulation.
\newblock In \emph{Algorithmic Learning Theory}, 89--103.

\bibitem[{de~Kerchove et~al.(2008)de~Kerchove, Ninove, and van
  Dooren}]{dekerchove08}
de~Kerchove, C., Ninove, N., and van Dooren, P. (2008).
\newblock Maximizing pagerank via outlinks.
\newblock \emph{Linear Algebra and its Applications}, 429(5), 1254 -- 1276.

\bibitem[{Friedkin and Johnsen(1990)}]{NEF-ECJ:90}
Friedkin, N.E. and Johnsen, E.C. (1990).
\newblock Social influence and opinions.
\newblock \emph{J. of Math. Sociology}, 15(3-4), 193--206.

\bibitem[{Galeotti et~al.(2017)Galeotti, Golub, , and Goyal}]{Galeotti17}
Galeotti, A., Golub, B., , and Goyal, S. (2017).
\newblock Targeting interventions in networks.
\newblock \urlprefix\url{https://arxiv.org/abs/1710.06026}.
\newblock Cambridge Working Papers in Economics 1744.

\bibitem[{Galeotti and Goyal(2009)}]{Galeotti09}
Galeotti, A. and Goyal, S. (2009).
\newblock Influencing the influencers: a theory of strategic diffusion.
\newblock \emph{RAND Journal of Economics}, 40(3), 509--532.

\bibitem[{Ishii and Tempo(2014)}]{Ishii.Tempo:2014}
Ishii, H. and Tempo, R. (2014).
\newblock The pagerank problem, multiagent consensus, and web aggregation: A
  systems and control viewpoint.
\newblock \emph{IEEE Control Systems Magazine}, 34(3), 34--53.

\bibitem[{Jackson(2005)}]{Jackson05}
Jackson, M.O. (2005).
\newblock \emph{A Survey of Models of Network Formation: Stability and
  Efficiency}, chapter~1, 11--57.

\bibitem[{Kempe et~al.(2015)Kempe, Kleinberg, and Tardos}]{Kempe15}
Kempe, D., Kleinberg, J., and Tardos, E. (2015).
\newblock Maximizing the spread of influence through a social network.
\newblock \emph{Theory of Computing}, 11(4), 105--147.

\bibitem[{Latora et~al.(2017)Latora, Nicosia, and Russo}]{latora17}
Latora, V., Nicosia, V., and Russo, G. (2017).
\newblock \emph{Complex Networks: Principles, Methods and Applications}.
\newblock Cambridge University Press.

\bibitem[{Monderer and Shapley(1996)}]{Monderer}
Monderer, D. and Shapley, L.S. (1996).
\newblock Potential games.
\newblock \emph{Games and Economic Behavior}, 14(1), 124 -- 143.

\bibitem[{Norris(1997)}]{norris_1997}
Norris, J.R. (1997).
\newblock \emph{Markov Chains}.
\newblock Cambridge Series in Statistical and Probabilistic Mathematics.
  Cambridge University Press.

\end{thebibliography}
                                                   







\end{document}